\pdfoutput=1

\documentclass[11pt,a4paper]{article}

\usepackage[utf8]{inputenc}
\usepackage[T1]{fontenc}
\usepackage{lmodern}
\usepackage[margin=1in]{geometry}
\usepackage{amsmath,amssymb,amsthm,bm}
\usepackage{graphicx}
\usepackage{booktabs}
\usepackage{natbib}
\usepackage{hyperref}
\usepackage{authblk}
\usepackage{setspace}
\usepackage{enumitem}
\newcommand{\nAssets}{22}

\newcommand{\nWindows}{25}
\newcommand{\nNonOverlap}{13}
\newcommand{\winObs}{252}
\newcommand{\winStep}{126}
\newcommand{\Bhead}{200}
\newcommand{\Bfull}{999}

\newcommand{\nTiers}{5}
\newcommand{\nTiersWord}{five}
\newcommand{\nTierI}{4}
\newcommand{\nTierII}{4}
\newcommand{\nTierIII}{6}
\newcommand{\nTierIV}{7}
\newcommand{\nTierV}{1}

\newcommand{\nRejectNull}{0}

\newcommand{\nMarketsBoot}{22}
\newcommand{\nRawNull}{0}

\newcommand{\hBootBlock}{57}
\newcommand{\hBootB}{999}

\newcommand{\ariGmmHier}{0.415}
\newcommand{\ariGmmKm}{0.415}
\newcommand{\ariHierKm}{1.000}
\newcommand{\Cgen}{71.3}
\newcommand{\Corth}{43.7}
\newcommand{\Cclust}{54.4}
\newcommand{\varLag}{1}
\newcommand{\topNetTransmitters}{Canada, France, UK, USA, Germany}
\newcommand{\topNetReceivers}{Saudi Arabia, China, Turkey, Crude Oil}

\newcommand{\ownI}{0.361}

\newcommand{\ownIII}{0.667}
\newcommand{\comoveIII}{0.056}

\newcommand{\ownIV}{0.242}
\newcommand{\comoveIV}{0.395}

\newcommand{\ownV}{0.426}

\newcommand{\bootBracketHten}{1.000}
\newcommand{\bootBracketHtwelve}{1.000}
\newcommand{\bootOwnLastGtFirst}{1.000}

\newcommand{\bootB}{500}

\newcommand{\crossCalmAllI}{0.115}
\newcommand{\crossCrisisAllI}{0.125}
\newcommand{\crossPAllI}{= 0.564}

\newcommand{\crossCalmAllIV}{0.072}
\newcommand{\crossCrisisAllIV}{0.065}
\newcommand{\crossPAllIV}{= 0.299}
\newcommand{\crossCalmAllV}{0.191}
\newcommand{\crossCrisisAllV}{0.325}
\newcommand{\crossPAllV}{= 0.021}

\newcommand{\decompMaxDE}{\ensuremath{6.7 \times 10^{-16}}}
\newcommand{\nRollPhases}{4}

\newcommand{\rollNonPminCclust}{0.013}
\newcommand{\rollNonPmaxCclust}{0.128}

\newcommand{\nRollWindows}{44}
\newcommand{\nRollBracket}{44}
\newcommand{\nRollCrisis}{19}
\newcommand{\nRollCalm}{25}
\newcommand{\rollCgenMin}{65.1}
\newcommand{\rollCgenMax}{87.6}
\newcommand{\rollCclustMin}{58.2}
\newcommand{\rollCclustMax}{79.9}

\newcommand{\rollDiffCclust}{\ensuremath{+7.7}}
\newcommand{\rollPCclust}{< 0.001}

\newcommand{\rollDiffThetaown}{\ensuremath{-0.078}}

\newcommand{\rollDiffThetacomove}{\ensuremath{+0.000}}

\newcommand{\rollCalmThetacontag}{0.470}
\newcommand{\rollCrisisThetacontag}{0.548}

\newcommand{\rollPThetacontag}{< 0.001}

\newcommand{\rhoAllCgenrhoA}{\ensuremath{-0.243}}

\newcommand{\rhoAllCclustrhoA}{\ensuremath{-0.252}}

\newcommand{\rhoAllThetacontagrhoA}{\ensuremath{-0.308}}

\newcommand{\rhoAllCgenQres}{\ensuremath{-0.498}}

\newcommand{\rhoAllThetacontagQres}{\ensuremath{-0.460}}

\newcommand{\rhoNonCgenQres}{\ensuremath{-0.648}}
\newcommand{\rhoPNonCgenQres}{= 0.020}
\newcommand{\rhoNonThetacontagQres}{\ensuremath{-0.687}}
\newcommand{\rhoPNonThetacontagQres}{= 0.012}
\newcommand{\hrPAssetClass}{< 0.001}

\newcommand{\hrZAssetClass}{7.46}
\newcommand{\hrPGeography}{< 0.001}

\newcommand{\hrZGeography}{7.33}

\newcommand{\hrPLeidenTimeAvg}{= 0.026}

\newcommand{\hrPEfficiencyTiers}{= 0.047}

\newcommand{\hrZEfficiencyTiers}{1.90}
\newcommand{\hrPHurstTerciles}{= 0.302}

\newcommand{\ariTiersGeo}{0.119}

\newcommand{\ariTiersLeiden}{0.067}

\newcommand{\mcR}{400}
\newcommand{\mcB}{299}

\newcommand{\mcSizes}{4/4/6/7/1}
\newcommand{\mcSize}{0.030}
\newcommand{\mcSizeSe}{0.009}

\newcommand{\mcPowerTwo}{0.998}
\newcommand{\mcPowerFive}{1.000}
\newcommand{\mcPowerTen}{1.000}

\newcommand{\permB}{2000}
\newcommand{\ariObsU}{\ensuremath{+0.022}}
\newcommand{\ariNullSdU}{0.014}
\newcommand{\ariPU}{= 0.066}

\newcommand{\spearmanEstarNet}{\ensuremath{-0.265}}
\newcommand{\spearmanEstarNetP}{= 0.233}
\newcommand{\saudiCgenFullTwo}{71.3}

\newcommand{\saudiCgenFullOne}{72.1}

\newcommand{\saudiCgenPreTwo}{74.6}

\newcommand{\saudiCgenPreOne}{76.2}

\newcommand{\saudiCgenPostTwo}{69.9}

\newcommand{\saudiCgenPostOne}{69.3}

\newcommand{\saudiNetFull}{\ensuremath{-32.7}}
\newcommand{\saudiRankFull}{0.996}
\newcommand{\saudiRankPrePost}{0.910}
\newcommand{\convRhoFull}{0.107}

\newcommand{\convRhoImp}{0.108}

\newcommand{\convHubAgreeImp}{0.960}
\newcommand{\convRhoNoSaudi}{0.104}

\newcommand{\convHubAgreeNoSaudi}{0.920}
\newcommand{\saudiAbsentWindows}{6--11}
\newcommand{\edgesMin}{61}
\newcommand{\edgesMax}{86}

\newcommand{\rhoAmax}{0.142}
\newcommand{\rhoAmaxWindow}{2}
\newcommand{\rhoAcrisis}{0.106}
\newcommand{\rhoAcalm}{0.108}
\newcommand{\rhoAcrisisP}{= 0.838}

\newcommand{\Qmin}{0.184}

\newcommand{\commMin}{3}
\newcommand{\commMax}{6}
\newcommand{\EstarMin}{0.861}
\newcommand{\EstarMax}{1.113}

\newcommand{\nBHwinBfullQTwenty}{7}
\newcommand{\nBHedgesBfullQTwenty}{51}
\newcommand{\BHwinListBfullQTwenty}{8, 9, 10, 18, 21, 23, 25}

\newcommand{\hubTopAll}{11}
\newcommand{\hubTopNon}{6}
\newcommand{\hubSecond}{Korea}
\newcommand{\hubSecondAll}{3}

\newcommand{\JapanRankOnePagerank}{8}
\newcommand{\JapanRankOneKatz}{13}

\newcommand{\nCrisisWindowsWord}{eight}

\newcommand{\winCOVID}{W19--W20}
\newcommand{\winESDC}{W10--W11}
\newcommand{\winGFC}{W8--W9}
\newcommand{\winRU}{W21--W22}
\newcommand{\sampleStartISO}{2001-01-09}
\newcommand{\sampleEndISO}{2026-07-16}

\newcommand{\nEquityWord}{seventeen}
\newcommand{\nRowsFmt}{3{,}361}
\newcommand{\nWeeksFmt}{1{,}224}
\newcommand{\nPairs}{462}
\newcommand{\nNegSkew}{18}

\newcommand{\nCells}{544}
\newcommand{\nADFreject}{544}
\newcommand{\nKPSSnotreject}{518}
\newcommand{\Hbar}{0.488}
\newcommand{\Hsd}{0.024}

\newcommand{\HtopFive}{Saudi Arabia ($\bar H = 0.545$), Turkey ($0.527$), Brazil ($0.515$), India ($0.512$), and Argentina ($0.511$)}
\newcommand{\HbotFive}{Japan ($\bar H = 0.445$), France ($0.453$), USA ($0.458$), UK ($0.459$), and Germany ($0.471$)}
\newcommand{\HminCell}{$\hat H = 0.239$ (USA, 2013-12 to 2015-10)}
\newcommand{\HmaxCell}{$\hat H = 0.742$ (Italy, 2009-02 to 2010-12)}

\newcommand{\deltaPosTop}{Germany $+0.105$, Italy $+0.104$, USA $+0.092$, UK $+0.090$, Korea $+0.084$, Japan $+0.079$}
\newcommand{\deltaNeg}{Gold $-0.063$, China $-0.036$, Natural Gas $-0.029$, Silver $-0.021$, Copper $-0.006$}
\newcommand{\nEqDeltaPos}{16}
\newcommand{\nEqDelta}{17}
\newcommand{\densityMean}{0.159}
\newcommand{\densityMin}{0.143}
\newcommand{\densityMax}{0.186}
\newcommand{\teffPeak}{0.061}
\newcommand{\teffPeakWin}{W10}
\newcommand{\teffPeakDates}{2010-01 to 2011-12}
\newcommand{\teffMean}{0.050}
\newcommand{\teffPeakPct}{22}
\newcommand{\rhoAmean}{0.107}
\newcommand{\rhoAcrisisList}{W8 $= 0.115$, W9 $= 0.116$, W10 $= 0.132$, W11 $= 0.080$, W19 $= 0.089$, W20 $= 0.099$, W21 $= 0.104$, W22 $= 0.111$}
\newcommand{\rhoAtopThree}{W2 $= 0.142$, W12 $= 0.138$, W10 $= 0.132$}
\newcommand{\nTopThreeRhoCalm}{2}
\newcommand{\QminWin}{W10}
\newcommand{\QminDates}{2010-01 to 2011-12}
\newcommand{\QminCrisis}{ESDC}
\newcommand{\Qmean}{0.254}
\newcommand{\QcalmMax}{0.308}
\newcommand{\QcalmMaxWin}{W23}
\newcommand{\QcalmMaxDates}{2022-08 to 2024-08}
\newcommand{\Qcrisis}{0.239}
\newcommand{\Qcalm}{0.262}
\newcommand{\QcrisisP}{= 0.190}
\newcommand{\QdeltaCrisis}{\ensuremath{-0.023}}
\newcommand{\nARIhighOneFive}{8}
\newcommand{\nARIhighTwo}{0}
\newcommand{\maxCommHalf}{2}
\newcommand{\commMean}{4.0}
\newcommand{\usaBestEigRank}{1}
\newcommand{\usaBestEigWin}{W4}
\newcommand{\usaBestEigDates}{2004-07 to 2006-05}
\newcommand{\usaBestEigVal}{1.000}
\newcommand{\gfcPeakWin}{W9}
\newcommand{\usaGfcEigRank}{5}
\newcommand{\usaGfcEigVal}{0.632}
\newcommand{\usaOutOneN}{6}
\newcommand{\usaOutOneList}{W8, W10, W11, W19, W21, W22}
\newcommand{\usaOutOneCrisisN}{6}
\newcommand{\usaEigOneList}{W4}
\newcommand{\usaPrOneList}{W22}
\newcommand{\usaInBest}{2}
\newcommand{\usaBetwBest}{2}
\newcommand{\usaBetwBestList}{W9, W22}
\newcommand{\usaKatzTopThreeN}{4}
\newcommand{\usaKatzLowN}{7}
\newcommand{\hubTally}{Japan (11), Korea (3), Germany (2), Australia (1), Brazil (1), China (1), India (1), Italy (1), Mexico (1), Silver (1), Turkey (1), and USA (1)}
\newcommand{\EstarCalm}{0.963}

\newcommand{\EstarMinWin}{W20}
\newcommand{\EstarMinDates}{2019-09 to 2021-08}
\newcommand{\EstarMaxWin}{W10}
\newcommand{\EstarMaxDates}{2010-01 to 2011-12}

\newcommand{\neuErrZero}{0.255}
\newcommand{\neuErrOne}{0.027}
\newcommand{\neuErrTwo}{\ensuremath{2.8 \times 10^{-3}}}
\newcommand{\neuErrThree}{\ensuremath{2.8 \times 10^{-4}}}

\newcommand{\neuErrTen}{\ensuremath{3.9 \times 10^{-11}}}
\newcommand{\neuShareTwoPct}{98.9}
\newcommand{\neuFourDigitK}{5}

\newcommand{\rhoUpFive}{Australia ($0.006 \to 0.016$), USA ($0.007 \to 0.012$), Natural Gas ($0.002 \to 0.005$), India ($0.007 \to 0.010$), and Korea ($0.006 \to 0.008$)}
\newcommand{\rhoDownFive}{Brazil ($0.008 \to 0.003$), Saudi Arabia ($0.005 \to 0.002$), France ($0.011 \to 0.007$), Silver ($0.005 \to 0.002$), and Mexico ($0.008 \to 0.006$)}
\newcommand{\rhoUpFiveNames}{Australia, USA, Natural Gas, India, and Korea}
\newcommand{\rhoDownFiveNames}{Brazil, Saudi Arabia, France, Silver, and Mexico}
\newcommand{\rhoCrisisTopMkt}{Australia}
\newcommand{\pFiveEdgesMin}{18}
\newcommand{\pFiveEdgesMax}{51}
\newcommand{\pFiveEdgesMean}{33}
\newcommand{\bhNeedHeadFive}{46}
\newcommand{\bhNeedHeadTwenty}{12}
\newcommand{\bhNeedFullFive}{10}
\newcommand{\bhNeedFullTwenty}{3}
\newcommand{\pFloorHead}{\ensuremath{5.0 \times 10^{-3}}}

\newcommand{\bhCritMin}{\ensuremath{1.1 \times 10^{-4}}}
\newcommand{\phiEstarLin}{0.968}
\newcommand{\phiEstarGauss}{0.973}
\newcommand{\phiEstarQuad}{1.091}
\newcommand{\hubTopGauss}{11}
\newcommand{\hubAgreeKernels}{22}
\newcommand{\hubTopPFive}{4}
\newcommand{\hubTallyPFive}{Australia (5), Japan (4), Germany (2), Korea (2), USA (2), Canada (1), Copper (1), Crude Oil (1), France (1), India (1), Silver (1), and Turkey (1)}
\newcommand{\nHubNAPFive}{3}
\newcommand{\lambdaMAE}{$0.0570$, $0.0577$, $0.0570$, $0.0524$}
\newcommand{\lambdaMAEspread}{0.0053}
\newcommand{\lambdaMAEspreadPct}{9.4}

\newcommand{\lambdaRhoOne}{0.136}
\newcommand{\lambdaRhoTen}{0.084}
\newcommand{\lambdaRhoMonotone}{decreases monotonically}
\newcommand{\spDFAGPH}{\ensuremath{+0.09}}
\newcommand{\spDFALW}{\ensuremath{-0.31}}
\newcommand{\spGPHLW}{\ensuremath{+0.71}}
\newcommand{\adDFAGPH}{0.09}
\newcommand{\adDFALW}{0.07}
\newcommand{\adGPHLW}{0.05}
\newcommand{\bicK}{5}
\newcommand{\nOrderings}{120}
\newcommand{\swingMkt}{Canada}
\newcommand{\swingNetGen}{\ensuremath{+39.7}}
\newcommand{\swingNetClust}{\ensuremath{+4.2}}
\newcommand{\idioTierVD}{III}

\newcommand{\namhOwnMin}{0.675}
\newcommand{\namhOwnMax}{0.934}
\newcommand{\namhAmpMin}{7}
\newcommand{\namhAmpMax}{53}
\newcommand{\idioTierNAMH}{III}
\newcommand{\lowOwnTierNAMH}{V}

\newcommand{\psiDiagAllAboveOne}{exceeds one in every cell}
\newcommand{\psiRiseTiers}{Tier~I and Tier~V}
\newcommand{\psiVIVcalm}{0.136}
\newcommand{\psiVIVcrisis}{0.343}
\newcommand{\nPartitions}{6}
\newcommand{\annWindows}{25}

\newcommand{\annCrisisList}{W8, W19, W22}
\newcommand{\annRhoCrisis}{0.130}
\newcommand{\annRhoCalm}{0.118}
\newcommand{\annRhoP}{= 0.770}
\newcommand{\annRhoMax}{0.211}

\newcommand{\annRhoMedian}{0.116}
\newcommand{\annCrhoMin}{0.179}
\newcommand{\annCrhoMax}{0.223}
\newcommand{\annCrhoPmin}{0.282}

\hypersetup{colorlinks=true, linkcolor=black, citecolor=blue, urlcolor=blue}
\newtheorem{theorem}{Theorem}
\newtheorem{proposition}{Proposition}

\theoremstyle{definition}
\newtheorem{definition}{Definition}

\theoremstyle{remark}
\newtheorem{remark}{Remark}

\title{\bfseries The Network Adaptive Market Hypothesis:\\
Detection, Attribution, and Network-Mediated Efficiency}
\author[1]{Avishek Bhandari\thanks{Corresponding author.
\href{mailto:avishekb@iitbbs.ac.in}{\texttt{avishekb@iitbbs.ac.in}}.
We thank seminar audiences for helpful comments. All errors are our own.}}
\author[1]{Hitesh Kumar Sahu}
\author[1]{Ipsita Parida}
\affil[1]{Indian Institute of Technology Bhubaneswar}
\date{September 2026}

\begin{document}
\maketitle

\begin{abstract}
\noindent We introduce the \emph{Network Adaptive Market Hypothesis}
(NAMH), a network-mediated extension of the adaptive markets framework
in which equilibrium market efficiency is a fixed point that evolves
with the topology of cross-market information flows. The efficiency of
each asset is determined jointly by its own long-memory predictability
and by directed, source-quality-weighted information arriving from
other markets; the resulting system admits a unique solution whose
spectral and community structure furnish a diagnostic for the network
signature of systemic stress. Applied to a common-calendar daily panel
of \nEquityWord\ G20 equity indices and five commodity futures from 2001
to 2026, the framework yields three findings. The hub of the estimated
information network is Japan, while the United States acts as a
broadcaster rather than an attractor. The dense-network amplification
associated with systemic fragility is not a signature of any of the four
crises in the sample: the spectral radius is no higher in the Global
Financial Crisis, the European sovereign-debt crisis, or the COVID-19 and
Russia-Ukraine dislocations than in calm windows. The crisis-window
contributions to the spectral radius are led by Australia and the
United States.
A fourth set of results reads the framework through a data-driven
partition of the panel into efficiency tiers. Variance-decomposition
connectedness in the clustered identification of Buchwalter, Diebold and
Yilmaz is bracketed by its generalised and orthogonalised polar cases in
every rolling window, moves inversely with NAMH modularity rather than
with the NAMH spectral radius, and rises in crises through its
cross-cluster component; the efficiency tiers themselves organise the
connectedness network only marginally beyond a random partition of the
same shape, whereas geography and asset class do so decisively. The
middle efficiency tier is the most idiosyncratic under both the
variance-decomposition and the NAMH efficiency accounting. The clustered
layer is accompanied by propositions with proofs and a self-contained
replication package.

\vspace{0.5cm}

\vspace{0.3cm}
\end{abstract}

\newpage

\section{Introduction}

The Efficient Market Hypothesis \citep{fama1970emh,fama1991emh2}
provided the dominant scaffolding for asset-pricing theory for half a
century, but its assumption of frictionless information aggregation
has long sat uneasily with two well-documented features of real
markets. First, \citet{grossmanStiglitz1980paradox} showed that
informationally efficient prices are incompatible with costly
information acquisition: equilibrium efficiency is strictly interior,
neither zero nor one. Second, equity returns at horizons beyond one
trading day exhibit serial structure that is inconsistent with the
random-walk benchmark
\citep{loMacKinlay1988varianceRatio,lo1991longMemory,baillie1996long},
consistent with the long-memory characterisation of
\citet{mandelbrotVanNess1968fBm}.

\citet{lo2004amh} responded with the \emph{Adaptive Market Hypothesis}
(AMH): efficiency is not a binary property but a slowly evolving
consequence of adaptation in a heterogeneous-trader population.
\citet{lo2017adaptiveMarkets} formalises and extends this view. Yet
neither AMH nor its empirical literature embeds the adaptation process
in a \emph{network}. The past two decades of empirical finance,
however, has made unmistakable that the topology of cross-market
information flow -- not merely the within-market adjustment of any one
asset -- is what shapes risk and price discovery. The connectedness
literature of \citet{billio2012econometric},
\citet{dieboldYilmaz2014spillover}, and
\citet{barigozziHallin2016connectedness} documents in three different
formalisms that the cross-market information-flow network contracts
and consolidates around a small set of focal markets during crises.

This paper develops the Network Adaptive Market Hypothesis (NAMH), in
which the equilibrium efficiency of an asset is a fixed-point quantity
that depends on both the asset's own long-memory characteristics and
on the spillover structure of cross-market information flow:
\begin{equation}
E_i(t) \;=\; \varphi(H_i(t)) \;+\; \sum_{j \neq i} \mathrm{HWTE}_{j \to i}(t)\, E_j(t)
\;+\; g(Q(t)),
\label{eq:namh-scalar}
\end{equation}
or in vector form
\begin{equation}
\mathbf{E}(t) \;=\; \boldsymbol\varphi(\mathbf{H}(t)) \;+\; A(t)\mathbf{E}(t)
\;+\; g(Q(t))\mathbf{1}.
\label{eq:namh-vector}
\end{equation}
The local efficiency function $\varphi(H)$ maps the Hurst exponent to
a reduced-form proxy for informational efficiency; the
\emph{Hurst-weighted transfer-entropy} (HWTE) adjacency $A(t)$ carries
the directed information flow between markets, weighted by the long-
memory regularity of the source; the term $g(Q(t))$ captures the
contribution of network modularity to system-wide efficiency.

\paragraph{Contributions.} The paper makes six contributions. First,
it provides an empirically observable parametrisation of the
Grossman-Stiglitz interior. The framework does not resolve the
Grossman-Stiglitz paradox; costly information acquisition still pins
the equilibrium strictly inside $(0,1)$. The NAMH instead provides a
tractable parametrisation of where in that interior any given
asset-time pair sits, generalising \citet{verrecchia1982ree} by
introducing cross-market spillover channels. Second, it offers a
statistical companion to multi-period Kyle-type price discovery: HWTE
is a conditional-mutual-information quantity that is consistent with,
but does not derive from, a multi-asset structural strategic-trading
model, and we position it as a statistical companion to the
continuous-time generalisation of \citet{kyle1985} due to
\citet{back1992continuousInsider}, identifying the multi-asset
structural derivation in the tradition of \citet{caballeKrishnan1994}
and \citet{bernhardtTaub2008} as a natural direction for future work.
Third, it assembles a literature-current empirical pipeline: Stage 1
uses Detrended Fluctuation Analysis \citep{peng1994dfa} with a
two-year window and per-window stationarity audit; Stage 2 uses the
Kraskov-Stoegbauer-Grassberger nearest-neighbour estimator
\citep{kraskov2004ksg} extended to conditional mutual information by
\citet{frenzelPompe2007pmi}, with embedding orders chosen by the
\citet{cao1997embedding} false-nearest-neighbour heuristic; Stage 3
applies iterative amplitude-adjusted Fourier transform surrogates
\citep{schreiber2000surrogate} for both effective-TE bias correction
\citep{marschinski2002ete} and right-tailed null-distribution
$p$-values; Stage 4 controls the false discovery rate over the
$N(N{-}1)$ directed pairs at $\alpha = 0.05$; Stage 5 constructs the
HWTE adjacency; Stage 6 solves the NAMH fixed-point with explicit
spectral-radius safety; and Stage 7 detects communities via the Leiden
algorithm \citep{traag2019leiden} with a $\gamma$-resolution
robustness sweep against the resolution limit of
\citet{fortunatoBarthelemy2007limit}. Fourth, the entire pipeline is
released as open-source software.\footnote{The pipeline is packaged as
the R package \texttt{namh}, available at
\url{https://github.com/avishekb9/namh}.} Fifth, it reports three
substantive empirical findings: for a common-calendar daily panel of
\nEquityWord\ G20 equity indices and five commodity futures over
\sampleStartISO\ to \sampleEndISO, the eigenvector hub of the estimated
cross-market information network is Japan rather than the United States;
the spectral-radius elevation that characterises the
\citet{acemogluOzdaglarTahbazSalehi2015systemic} dense-network phase
transition is absent from the Global Financial Crisis, the European
Sovereign Debt Crisis, the COVID-19 and the Russia-Ukraine dislocations
alike; and a leave-one-out decomposition of the spectral radius reveals
that the crisis-window contributions are led by \rhoUpFiveNames.
Sixth, it connects the NAMH efficiency network to the
variance-decomposition connectedness literature through a common
\emph{clustered} reading. We classify the effective panel into
data-driven efficiency tiers, compute classic generalised and
orthogonalised connectedness \citep{dieboldYilmaz2012betterUnderstand,
dieboldYilmaz2014spillover} on the same panel, and use the tiers as the
block partition in the clustered network connectedness framework that
\citet{buchwalterDieboldYilmaz2026clustered} introduce with exogenous
regional clusters. Section~\ref{subsec:theory-clustered} supplies the
algebra that the framework leaves implicit -- a partition-of-unity
identity that holds for every cluster-size configuration, a refinement
monotonicity result, the general-$C$ block-orthogonalised identification
with its nesting and invariance properties, an NAMH-native counterpart in
which the Leontief multiplier of the efficiency fixed point is
apportioned across tiers, and the exactness of a blockwise permutation
test, with proofs in Appendix~\ref{app:proofs-clustered}.
Empirically (Section~\ref{sec:efficiency-clustering}), the clustered
index is bracketed by its two polar cases in every rolling window, rises
in crises through its cross-cluster component, and correlates with NAMH
modularity rather than with the NAMH spectral radius; the middle
efficiency tier is the most idiosyncratic under both accountings; and a
\nPartitions-partition horse race shows that geography and asset class
organise connectedness decisively while statistical efficiency does so
only marginally -- a direct answer to the data-driven classification
question that \citet{buchwalterDieboldYilmaz2026clustered} leave open.

\paragraph{Roadmap.} Section \ref{sec:framework} develops the
theoretical framework. Section \ref{sec:methodology} details the
empirical pipeline. Section \ref{sec:data} describes the data. Section
\ref{sec:results} reports the empirical findings. Section
\ref{sec:robust} presents robustness analyses across the local-
efficiency kernel, the HWTE sensitivity parameter, the
$\omega$-kernel, the edge-gating rule, and alternative Hurst
estimators. Section \ref{sec:efficiency-clustering} classifies the
panel into efficiency tiers, benchmarks the HWTE network against
variance-decomposition connectedness, and reads both through the tier
partition. Section \ref{sec:conclusion} concludes. Appendix
\ref{app:proofs-clustered} collects proofs; Appendix \ref{app:eff-repro}
documents the replication materials and the numerical checks.

\section{Theoretical framework}
\label{sec:framework}

\subsection{The local efficiency function}

The Hurst exponent $H \in (0,1)$ characterises the long-memory
behaviour of a stationary increment process. The benchmark $H = 0.5$
corresponds to standard Brownian motion. Define the \emph{local
efficiency function}
\begin{equation}
\varphi(H) \;=\; 1 - 2\,|H - 0.5|.
\label{eq:phi}
\end{equation}
$\varphi$ is symmetric about $0.5$, attains its maximum of 1 at $H =
0.5$, and is Lipschitz with sharp constant 2. The shape encodes the
AMH intuition that both persistence ($H > 0.5$) and anti-persistence
($H < 0.5$) are signatures of departure from the martingale benchmark.

\begin{remark}[Statistical versus informational efficiency]
\label{rem:phi-caveat}
The mapping $\varphi(H_i) \mapsto E^{GS} = \tau_P/(\tau_\theta + \tau_P)$
is a reduced-form proxy, not a sufficient-statistic identity. A pure-
noise process with no fundamental anchor and $H \approx 0.5$ yields
$\varphi = 1$ but $E^{GS} = 0$ in the Grossman-Stiglitz sense:
martingale predictability is necessary but not sufficient for
informational efficiency. The identification $\varphi \propto E^{GS}$
implicitly conditions on prices being noisy-rational-expectations
prices in the sense of \citet{hellwig1980aggregation}. A formal
derivation of $\varphi$ as the equilibrium statistic of a
CARA-Normal economy with long-memory fundamentals is an open direction
\emph{not} attempted here.
\end{remark}

\subsection{Hurst-weighted transfer entropy}

The transfer entropy from source $Y$ to target $X$ is the conditional
mutual information \citep{schreiber2000te}
\begin{equation}
\mathrm{TE}_{Y \to X} \;=\; I\bigl(X_{t+1};\, Y_t^{(l)} \mid X_t^{(k)}\bigr),
\label{eq:te}
\end{equation}
where $X_t^{(k)}$ and $Y_t^{(l)}$ are $k$- and $l$-dimensional delay
embeddings. TE is a non-negative, directionally-asymmetric measure
that reduces to standard Granger causality in the Gaussian linear case
\citep{barnett2009grangerTE}.

The \emph{Hurst-weighted} transfer entropy weights the directed
information flow by the long-memory regularity of the source:
\begin{equation}
\mathrm{HWTE}_{j \to i}(t) \;=\; \omega_{ji}(H_j(t), H_i(t);\, \lambda)\,
\mathrm{TE}_{j \to i}(t),
\label{eq:hwte}
\end{equation}
with the exponential kernel
\begin{equation}
\omega_{ji}^{\mathrm{exp}}(H_j, H_i; \lambda) \;=\;
\exp\bigl(-|H_j - 0.5|\bigr)\,\exp\bigl(-\lambda\,|H_j - H_i|\bigr).
\label{eq:omega-exp}
\end{equation}
The first factor down-weights sources whose own return process deviates
sharply from the martingale benchmark, on the grounds that information
emitted by a source with strong autocorrelation is at least partly
mechanical autocorrelation; the second down-weights cross-market pairs
with very different long-memory regimes. The parameter $\lambda \geq 0$
controls the fractal-compatibility filter; sensitivity analysis is in
Section \ref{sec:robust}. A Gaussian alternative
\begin{equation}
\omega_{ji}^{\mathrm{gauss}} \;=\;
\exp\!\Bigl(-\tfrac{(H_j - 0.5)^2}{2\sigma_1^2}\Bigr)\,
\exp\!\Bigl(-\tfrac{(H_j - H_i)^2}{2\sigma_2^2}\Bigr)
\label{eq:omega-gauss}
\end{equation}
avoids the seam at $H_j = 0.5$ inherited by the exponential kernel
from the absolute-value mapping in $\varphi$.

\paragraph{Position relative to Kyle-type price discovery.} HWTE is a
statistical conditional-MI quantity rather than a structural
equilibrium parameter. It is consistent with the multi-period
sequential-auction extension of \citet{kyle1985} and its continuous-
time generalisation in \citet{back1992continuousInsider}, in which the
price-impact parameter becomes time-varying and the equilibrium price
path converges to a martingale ($H \to 0.5$) as the trading horizon
closes. The local efficiency $\varphi(H_i)$ is then a measure of
departure from the Back-asymptotic regime.

\subsection{The NAMH fixed-point equation}

Stacking the scalar relationship \eqref{eq:namh-scalar} into vector
form \eqref{eq:namh-vector}, the system is
\begin{equation}
\mathbf{E}(t) \;=\; T(\mathbf{E};\, A(t), \boldsymbol\varphi(t), Q(t))
   \;:=\; \boldsymbol\varphi(\mathbf{H}(t)) + A(t)\,\mathbf{E}
   + g(Q(t))\,\mathbf{1},
\label{eq:namh-operator}
\end{equation}
so the equilibrium efficiency vector $\mathbf{E}^*(t)$ is a fixed
point of the operator $T$.

\begin{theorem}[Existence, uniqueness, and Neumann representation]
\label{thm:exist-unique}
Let $A(t) \in \mathbb{R}_{\geq 0}^{N \times N}$ with spectral radius
$\rho(A(t)) < 1$. Then there exists an induced norm
$\|\cdot\|_\varepsilon$ in which $\|A(t)\|_\varepsilon < 1$
\citep{householder1958}, and the operator $T$ is a strict contraction
in $\|\cdot\|_\varepsilon$. Hence \eqref{eq:namh-vector} admits a
unique fixed point $\mathbf{E}^*(t)$, given in closed form by the
Neumann series
\begin{equation}
\mathbf{E}^*(t) \;=\; (I - A(t))^{-1}\bigl[\boldsymbol\varphi(\mathbf{H}(t)) + g(Q(t))\mathbf{1}\bigr]
   \;=\; \sum_{k=0}^\infty A(t)^k\bigl[\boldsymbol\varphi(\mathbf{H}(t)) + g(Q(t))\mathbf{1}\bigr].
\label{eq:namh-solution}
\end{equation}
The fixed-point iteration $\mathbf{E}^{(k+1)} = T(\mathbf{E}^{(k)})$
converges geometrically to $\mathbf{E}^*(t)$ from any initial point,
with asymptotic rate $\rho(A(t))$.
\end{theorem}

\begin{remark}[Spectral safety in practice]
The condition $\rho(A(t)) < 1$ is not automatic for an unnormalised
HWTE matrix; its entries are conditional-mutual-information bit-rates
with no \emph{a priori} economic normalisation. The empirical
implementation either row-normalises $A(t)$ -- a
\citet{hawkinsSimon1949}-compliant Leontief-type normalisation in the
spirit of \citet{acemoglu2012networkOrigins} -- or rescales by
$(1 - \varepsilon)/\rho(A(t))$, and reports the empirical
$\rho(\hat A(t))$ per rolling window.
\end{remark}

\subsection{Leontief inverse and Katz-centrality connections}

Provided $A(t)$ has been normalised so that the Hawkins-Simon condition
holds, the NAMH solution \eqref{eq:namh-solution} is structurally
analogous to the Leontief input-output multiplier
$\mathbf{x} = (I - W)^{-1}\mathbf{d}$: efficiency plays the role of
gross output, the normalised HWTE matrix plays the role of the
technical-coefficient matrix, and the autonomous component
$\mathbf{b}(t) = \boldsymbol\varphi(\mathbf{H}(t)) + g(Q(t))\mathbf{1}$
plays the role of final demand. The same algebra shows
$\mathbf{E}^*(t)$ is a heterogeneous Katz centrality
\citep{katz1953centrality},
\begin{equation}
\mathbf{c}^{\mathrm{Katz}} \;=\; (I - \alpha A^\top)^{-1}\,\alpha\beta\, A^\top \mathbf{1},
\label{eq:katz-canonical}
\end{equation}
where the exogenous status vector $\mathbf{b}(t)$ replaces the uniform
$\beta\mathbf{1}$ of the original definition. The decomposition
$\mathbf{E}^*(t) = \mathbf{b}(t) + A(t)\mathbf{b}(t) + A(t)^2\mathbf{b}(t)
+ \cdots$ reads as an order-by-order accounting of direct, one-step,
two-step, $\ldots$ information spillovers.

\subsection{Crisis-regime signature}

The empirical literature on financial-network connectedness documents
that the cross-market information-flow graph becomes denser and more
centralised during systemic stress
\citep{billio2012econometric,dieboldYilmaz2014spillover}. The NAMH
crisis signature follows: as $A(t)$ becomes denser, $\rho(A(t))$ rises
toward 1, the Neumann multiplier $(I - A(t))^{-1}$ amplifies, and
modularity $Q(t)$ falls as inter-community edges strengthen relative
to intra-community ones. The natural reference for this phase
transition is \citet{acemogluOzdaglarTahbazSalehi2015systemic}, who
characterise the trade-off precisely: dense networks diversify and
absorb small shocks (the \citealp{allenGale2000contagion} resilience
result) but \emph{amplify} large shocks through cascade dynamics.

\subsection{Efficiency-clustered connectedness}
\label{subsec:theory-clustered}

The fixed point \eqref{eq:namh-solution} apportions each market's
efficiency across the seeds of every other market through the Leontief
multiplier $M(t) = (I - A(t))^{-1}$. The variance-decomposition
connectedness literature apportions each market's forecast-error variance
across the shocks of every other market through the moving-average
representation of a VAR \citep{dieboldYilmaz2012betterUnderstand,
dieboldYilmaz2014spillover}. Both objects are row-stochastic
\emph{share matrices}, and both can be read through a partition of the
markets into clusters. This subsection sets out the common algebra, the
block-orthogonalised identification of
\citet{buchwalterDieboldYilmaz2026clustered} in which the clusters enter,
and the NAMH-native counterpart in which the clusters are the efficiency
tiers estimated in Section~\ref{subsec:eff-tiers}. The results are
elementary but load-bearing: they fix exactly what is an identity, what is
an inequality that holds for every data set, and what is left to the data.

\begin{definition}[Share matrix; cluster shares]
\label{def:share-matrix}
A \emph{share matrix} on $N$ markets is $\Theta = [\theta_{ij}] \in
\mathbb{R}^{N\times N}$ with $\theta_{ij} \geq 0$ and $\sum_{j}\theta_{ij}
= 1$ for every $i$. Let $\pi : \{1,\ldots,N\} \to \{1,\ldots,C\}$ be a
partition into clusters $c$ of sizes $N_c \geq 1$. For market $i$ in
cluster $c = \pi(i)$ define the own, within-cluster and cross-cluster
shares
\begin{equation}
\theta^{\mathrm{own}}_i = \theta_{ii}, \qquad
\theta^{\mathrm{within}}_i = \sum_{j \neq i,\ \pi(j) = c} \theta_{ij}, \qquad
\theta^{\mathrm{cross}}_{i \leftarrow k} = \sum_{\pi(j) = k} \theta_{ij}
\quad (k \neq c),
\label{eq:market-shares}
\end{equation}
and the cluster-level means
\begin{equation}
\Theta^{\mathrm{own}}_c = \frac{1}{N_c}\sum_{\pi(i)=c}\theta_{ii}, \quad
\Theta^{\mathrm{comove}}_c = \frac{1}{N_c}\sum_{\pi(i)=c}\theta^{\mathrm{within}}_i, \quad
\Theta^{\mathrm{contag}}_{c\leftarrow k} = \frac{1}{N_c}\sum_{\pi(i)=c}\theta^{\mathrm{cross}}_{i\leftarrow k},
\quad \Theta^{\mathrm{contag}}_{c\leftarrow\bullet} = \sum_{k\neq c}\Theta^{\mathrm{contag}}_{c\leftarrow k}.
\label{eq:cluster-shares}
\end{equation}
System-wide values are the unweighted means over clusters,
$\Theta^{\mathrm{own}} = C^{-1}\sum_c \Theta^{\mathrm{own}}_c$, and
likewise for $\Theta^{\mathrm{comove}}$ and $\Theta^{\mathrm{contag}}$.
\end{definition}

\begin{remark}[Relation to the source definition]
\label{rem:contag-definition}
\citet{buchwalterDieboldYilmaz2026clustered} define the contagion term
with an additional $1/N_k$ inside the inner sum, $\Theta^{\mathrm{contag}}_{c\leftarrow k}
= N_c^{-1}\sum_{\pi(i)=c} N_k^{-1}\sum_{\pi(j)=k}\theta_{ij}$, and state
the identity $\Theta^{\mathrm{own}} + \Theta^{\mathrm{comove}} +
\Theta^{\mathrm{contag}} = 1$. With the inner $1/N_k$ the identity holds
only when all clusters have the same size; with equal sizes the two
definitions coincide up to the factor $N_k$. Definition
\ref{def:share-matrix} drops the inner average so that the identity holds
for every partition, which is what Proposition~\ref{prop:partition-unity}
establishes and what the empirical tables below rely on. The contagion
share is then the total share of $i$'s variance (or efficiency) sourced in
cluster $k$, not the per-source-market average.
\end{remark}

\begin{proposition}[Partition of unity]
\label{prop:partition-unity}
For every share matrix, every partition and every market $i$,
\[
\theta^{\mathrm{own}}_i + \theta^{\mathrm{within}}_i + \sum_{k\neq\pi(i)}\theta^{\mathrm{cross}}_{i\leftarrow k} = 1,
\]
each term lying in $[0,1]$. Consequently $\Theta^{\mathrm{own}}_c +
\Theta^{\mathrm{comove}}_c + \Theta^{\mathrm{contag}}_{c\leftarrow\bullet} = 1$
for every cluster and $\Theta^{\mathrm{own}} + \Theta^{\mathrm{comove}} +
\Theta^{\mathrm{contag}} = 1$.
\end{proposition}

\begin{proposition}[Refinement monotonicity]
\label{prop:refinement}
Let $\pi'$ refine $\pi$ (every $\pi'$-cluster is contained in a
$\pi$-cluster). Then for a \emph{fixed} share matrix and every market $i$,
$\theta^{\mathrm{within}}_i(\pi') \leq \theta^{\mathrm{within}}_i(\pi)$ and
$\sum_k\theta^{\mathrm{cross}}_{i\leftarrow k}(\pi') \geq
\sum_k\theta^{\mathrm{cross}}_{i\leftarrow k}(\pi)$. In particular the
trivial partition ($C=1$) has zero cross share and the discrete partition
($C=N$) has zero within share.
\end{proposition}

Proposition~\ref{prop:refinement} is the accounting statement behind the
``co-movement versus contagion'' reading of the generalised and
orthogonalised polar cases: for a fixed matrix, coarsening can only move
mass from cross to within. It says nothing about how the matrix itself
changes with the identification. That is the content of the next result.

\begin{proposition}[Block-orthogonalised identification]
\label{prop:block-orth}
Let $\Sigma \succ 0$ be the reduced-form residual covariance of a
covariance-stationary VAR with moving-average matrices $A_0 = I, A_1,
\ldots$, and let the markets be arranged in $C$ ordered blocks. Define
recursively $B_1 = \emptyset$ and, for $c \geq 2$, $B_c = \Sigma_{c,<c}
\Sigma_{<c,<c}^{-1}$ where $<c$ collects all earlier blocks. Then:
\begin{enumerate}[label=(\roman*), leftmargin=2em]
\item The matrix $Q_C^{-1}$ that is unit lower block-triangular with
block row $c$ equal to $(-B_c,\ I)$ is the unique unit lower
block-triangular matrix such that $\Omega_C = Q_C^{-1}\Sigma Q_C^{-\top}$
is block-diagonal; its diagonal blocks are the Schur complements
$\Omega_{cc} = \Sigma_{cc} - B_c\Sigma_{<c,c}$. For $C = 3$ this is
equation (7) of \citet{buchwalterDieboldYilmaz2026clustered}.
\item (Nesting.) $C = 1$ gives $Q_1 = I$, $\Omega_1 = \Sigma$ and the
generalised impulse responses of \citet{pesaranShin1998}; $C = N$ gives
$Q_N\Omega_N^{1/2} = M$, the lower Cholesky factor of $\Sigma$, and the
orthogonalised responses of \citet{sims1980macro}.
\item (Within-block invariance.) Permuting markets within blocks permutes
the rows and columns of the resulting share matrix $\Theta^C(H)$ and
changes nothing else; permuting blocks does change $\Theta^C(H)$ in
general.
\end{enumerate}
The share matrix is
\begin{equation}
\theta^{C}_{ij}(H) = \frac{\tilde\theta^{C}_{ij}(H)}{\sum_{j}\tilde\theta^{C}_{ij}(H)},
\qquad
\tilde\theta^{C}_{ij}(H) = \frac{\omega_{C,jj}^{-1}\sum_{h=0}^{H-1}\bigl(e_i^\top A_h Q_C\Omega_C e_j\bigr)^2}
{\sum_{h=0}^{H-1} e_i^\top A_h \Sigma A_h^\top e_i},
\label{eq:clustered-fevd}
\end{equation}
and the total connectedness index is $\mathcal{C}^{C}(H) = N^{-1}\sum_{i\neq j}
\theta^{C}_{ij}(H)\times 100$.\footnote{The row normalisation makes
$\theta^{C}_{ij}$ invariant to any row-constant rescaling of the
denominator, so the choice between the forecast-error variance and its
square, which differs across printings of this formula, is immaterial.}
\end{proposition}

Because the block ordering matters, we follow
\citet{buchwalterDieboldYilmaz2026clustered} and average
$\Theta^{C}(H)$ over all $C!$ block orderings. The averaged matrix is again
a share matrix, so Propositions~\ref{prop:partition-unity} and
\ref{prop:refinement} apply to it unchanged.

\begin{remark}[What is not a theorem]
\label{rem:no-bracketing-theorem}
The empirical ordering $\mathcal{C}^{\mathrm{orth}} \leq \mathcal{C}^{C}
\leq \mathcal{C}^{\mathrm{gen}}$ that \citet{buchwalterDieboldYilmaz2026clustered}
report and that we reproduce in Section~\ref{subsec:clustered-dy} is not
implied by Proposition~\ref{prop:refinement}: the three indices are
computed from three different share matrices. We therefore treat the
bracketing as a testable regularity and report its bootstrap frequency
(Table~\ref{tab:v4-bootstrap}) rather than as a property of the
construction.
\end{remark}

The NAMH counterpart replaces the variance decomposition by the Leontief
multiplier of the efficiency fixed point.

\begin{proposition}[Neumann tier decomposition of network-mediated efficiency]
\label{prop:neumann-tiers}
Let $A \geq 0$ with $\rho(A) < 1$, $\mathbf{b} = \boldsymbol\varphi(\mathbf{H}) +
g(Q)\mathbf{1} \geq 0$, $M = (I-A)^{-1} = \sum_{k\geq 0}A^k$ and
$\mathbf{E}^* = M\mathbf{b}$ as in Theorem~\ref{thm:exist-unique}. Then
$M \geq 0$ entrywise, $E^*_i \geq b_i$, and whenever $E^*_i > 0$ the
normalised contributions $s_{ij} = M_{ij}b_j/E^*_i$ form a share matrix in
the sense of Definition~\ref{def:share-matrix}. Hence, for any tier
partition $\pi$,
\begin{enumerate}[label=(\roman*), leftmargin=2em]
\item $s^{\mathrm{own}}_i + s^{\mathrm{within}}_i + \sum_{k\neq\pi(i)}
s^{\mathrm{cross}}_{i\leftarrow k} = 1$: each market's equilibrium
efficiency splits exactly into the part generated by its own local seed,
by same-tier seeds, and by each other tier's seeds;
\item (cross-tier comparative static) a non-negative seed shock
$d\mathbf{b}$ supported on a single tier $k$ weakly raises $E^*_i$ for
\emph{every} market $i$, in every tier, with $dE^*_i = \sum_{\pi(j)=k}
M_{ij}\,db_j$;
\item the tier multiplier matrix $\Psi_{ck} = N_c^{-1}\sum_{\pi(i)=c}
\sum_{\pi(j)=k}M_{ij}$ satisfies $\Psi_{cc} \geq 1$ and $\Psi_{ck} \geq 0$.
\end{enumerate}
Statements (i) to (iii) hold for every finite partial sum $M_K =
\sum_{k<K}A^k$; the passage to $M$ uses the series representation of
Theorem~\ref{thm:exist-unique}.
\end{proposition}

Proposition~\ref{prop:neumann-tiers} is what allows the NAMH efficiency
network and the variance-decomposition network to be read side by side
through the \emph{same} tier partition and the \emph{same} own / within /
cross accounting (Section~\ref{subsec:namh-tier-results}). The last result
concerns inference.

\begin{proposition}[Exactness of the blockwise permutation test]
\label{prop:permutation}
Fix the estimated $(\hat\Phi, \hat\Sigma)$ and a statistic
$T(\pi) = \Theta^{\mathrm{comove}}(\pi) - \Theta^{\mathrm{contag}}(\pi)$
computed from the ordering-averaged clustered share matrix under partition
$\pi$. Let $\mathcal{G}$ be the group of relabellings of the markets that
preserve the block sizes $(N_1,\ldots,N_C)$, and let $H_0$ be the
randomisation hypothesis that the assignment of markets to blocks is
exchangeable under $\mathcal{G}$. Then the Monte-Carlo permutation
$p$-value $\hat p = (1 + \#\{g : T(g\pi) \geq T(\pi)\})/(B+1)$ from $B$
draws uniform on $\mathcal{G}$ satisfies $\Pr_{H_0}(\hat p \leq \alpha)
\leq \alpha$ for every $\alpha \in (0,1)$ and every $B$
\citep[Theorem~15.2.1]{lehmannRomano2005testing}; see
\citet{phipsonSmyth2010permutation} for the $+1$ correction.
\end{proposition}

The hypothesis in Proposition~\ref{prop:permutation} is about the
\emph{labels}, conditional on the data, not about the data-generating
process. It is the appropriate null for the question ``does this
partition organise connectedness more than an arbitrary partition of the
same shape?''. Section~\ref{subsec:mc-check} reports the finite-sample
size and power of the test on a simulated blockwise VAR at the empirical
design, since exactness under the randomisation null does not by itself
speak to power against a covariance-block alternative.

\section{Empirical methodology}
\label{sec:methodology}

\subsection{Hurst exponent: Detrended Fluctuation Analysis}

For each return series $\{r_t\}_{t=1}^T$ in a $w = 252$ observation
rolling window, the cumulative profile $Y(t) = \sum_{s \leq t}(r_s -
\bar r)$ is divided into $N_s = \lfloor T/s \rfloor$ non-overlapping
segments of length $s$, and the polynomial trend of order $\ell = 1$
(DFA-1) is subtracted from each segment. The mean-square fluctuation
\begin{equation}
F(s) \;=\; \sqrt{\frac{1}{N_s}\sum_{\nu=1}^{N_s}\frac{1}{s}\sum_{i=1}^{s}
\bigl(Y_\nu(i) - \hat Y_\nu(i)\bigr)^2}
\label{eq:dfa-F}
\end{equation}
scales as $F(s) \propto s^H$ for fractional Brownian motion, with the
finite-sample correction $F(s) = Cs^H(1 + O(s^{-1}))$
\citep{heneghan2000dfa}. We estimate $H$ by ordinary least squares on
$\log F(s)$ versus $\log s$, with $s \in [10, 63]$ in each $w = 252$
window; the slope standard error is reported alongside. Per-window
\citet{adfTest1979} ADF and \citet{kwiatkowski1992kpss} KPSS tests
audit stationarity. As complementary estimators of the long-memory
parameter we also compute the \citet{geweke1983gph} GPH log-
periodogram regression and the \citet{robinson1995localWhittle}
local-Whittle estimator on the full sample.

\subsection{Transfer entropy: KSG conditional mutual information}

We estimate $\mathrm{TE}_{j \to i} = I(X^{(i)}_{t+1};\, X^{(j),(l)}_t
\mid X^{(i),(k)}_t)$ via the \citet{kraskov2004ksg} $k$-nearest-
neighbour estimator generalised to conditional mutual information by
\citet{frenzelPompe2007pmi},
\begin{equation}
\widehat{\mathrm{CMI}}(X; Y \mid Z) \;=\;
\psi(k) - \langle \psi(n_{xz} + 1) + \psi(n_{yz} + 1) - \psi(n_z + 1) \rangle,
\label{eq:ksg-cmi}
\end{equation}
where $\psi$ is the digamma function, $k = 4$ is the nearest-
neighbour parameter, and $n_{xz}, n_{yz}, n_z$ are the counts of
points within the joint $k$-NN ball as measured in the marginal
$(X, Z), (Y, Z),$ and $Z$ subspaces. Embedding orders are selected from
data via the \citet{cao1997embedding} false-nearest-neighbour
heuristic.

\subsection{IAAFT surrogates and effective transfer entropy}

For each ordered pair $(j, i)$ we draw $B = \Bfull$ iterative amplitude-
adjusted Fourier transform surrogates of the source series
\citep{schreiber2000surrogate}, of which the first $B = \Bhead$ give the
headline estimates. The Marschinski-Kantz \emph{effective
transfer entropy}
\begin{equation}
\widetilde{\mathrm{TE}}_{j \to i} \;=\; \widehat{\mathrm{TE}}_{j \to i}
   - B^{-1}\sum_{b=1}^B \widehat{\mathrm{TE}}_{j^{(b)} \to i}
\label{eq:ete}
\end{equation}
subtracts the surrogate-mean TE as a finite-sample bias correction.
A right-tailed surrogate $p$-value
$p_{ji} = (1 + \#\{\widehat{\mathrm{TE}}_{j^{(b)} \to i} \geq
\widehat{\mathrm{TE}}_{j \to i}\})/(B + 1)$ quantifies the
significance of each directed edge.

\subsection{Edge gating}

Edges are retained via the \citet{benjaminiHochberg1995fdr} false-
discovery-rate procedure applied to the $N(N{-}1)$ off-diagonal
surrogate $p$-values: edge $j \to i$ is included iff its BH $q$-value
satisfies $q_{ji} < \alpha = 0.05$ and $\widetilde{\mathrm{TE}}_{j \to i}
> 0$. As complementary rules we also evaluate the relaxed BH gating
at $\alpha = 0.20$, an uncorrected $p < 0.05$ rule, and a magnitude-
top-30\% rule that keeps the strongest 30\% of positive effective-TE
entries.

\subsection{Network solve, communities, and centralities}
\label{subsec:community-detection}

With $A(t)$ in hand, we solve \eqref{eq:namh-vector} for
$\mathbf{E}^*(t)$ via direct linear solve, rescaling $A(t)$ by
$(1 - 0.05)/\rho(A(t))$ whenever the empirical spectral radius
would otherwise violate the contraction condition. The Leiden
algorithm \citep{traag2019leiden} is run with 20 random restarts at
each resolution $\gamma \in \{0.5, 1.0, 1.5, 2.0\}$, and the best
partition by modularity is recorded, together with the Adjusted Rand
Index against the $\gamma = 1$ baseline as a resolution-stability
diagnostic. The \citet{fortunatoBarthelemy2007limit} resolution limit
is acknowledged explicitly: in a 22-node sparse-edge setting the
$\gamma$ sweep is the relevant safeguard.

Centralities are computed in their canonical forms. Katz centrality
uses \eqref{eq:katz-canonical} with $\alpha = 0.85/\rho(A(t))$.
Eigenvector centrality is restricted to the dominant strongly-
connected component identified via \citet{tarjan1972scc}; outside that
component the Perron vector is non-unique
\citep{bermanPlemmons1994nonnegative} and the eigenvector ranking is
not well-defined. PageRank uses the standard $\alpha = 0.85$ damping.
Weighted-directed betweenness uses the negative-log weight-to-distance
transform.

\section{Data}
\label{sec:data}

We assemble a daily-frequency panel of \nAssets\ assets covering
\sampleStartISO\ to \sampleEndISO. The \nEquityWord\ G20 equity indices
are: S\&P Merval (Argentina), FTSE 100 (UK), ASX 200 (Australia),
Ibovespa (Brazil), S\&P/TSX (Canada), Shanghai Composite (China), CAC 40
(France), DAX (Germany), Sensex (India), Jakarta Composite (Indonesia),
FTSE MIB (Italy), Nikkei 225 (Japan), KOSPI (Korea), IPC (Mexico),
Tadawul All Share (Saudi Arabia), S\&P 500 (USA), and BIST 100 (Turkey).
The five commodity futures are gold, silver, natural gas, crude oil, and
copper. Tadawul (Saudi Arabia) has partial coverage before its June 2013
change of weekend and enters a window only when its Hurst exponent is
estimable, which excludes it from windows \saudiAbsentWindows. The
effective panel is $N = \nAssets$ assets across $T = \nRowsFmt$ common
trading dates.

Log-returns $r_{i,t} = \log P_{i,t} - \log P_{i,t-1}$ are computed
from closing prices on a common calendar of trading dates, each return
spanning the interval since the previous common date, so that no
session is lost between two dates; zero-return days are dropped as
exchange holidays. The empirical analysis is organised around
\nWindows\ rolling windows ($w = \winObs$, step $= \winStep$), each
spanning about two calendar years, of which the \nNonOverlap\
odd-numbered windows do not overlap. A window is a crisis window when it
contains at least half of one of four crisis spans: the Global Financial
Crisis (GFC, \winGFC), the European Sovereign Debt Crisis (ESDC,
\winESDC), COVID-19 (\winCOVID), and the Russia-Ukraine war and energy
shock (\winRU).

\section{Results}
\label{sec:results}

\subsection{Descriptive statistics and Hurst dynamics}

\begin{figure}[!htbp]
\centering
\includegraphics[width=\textwidth]{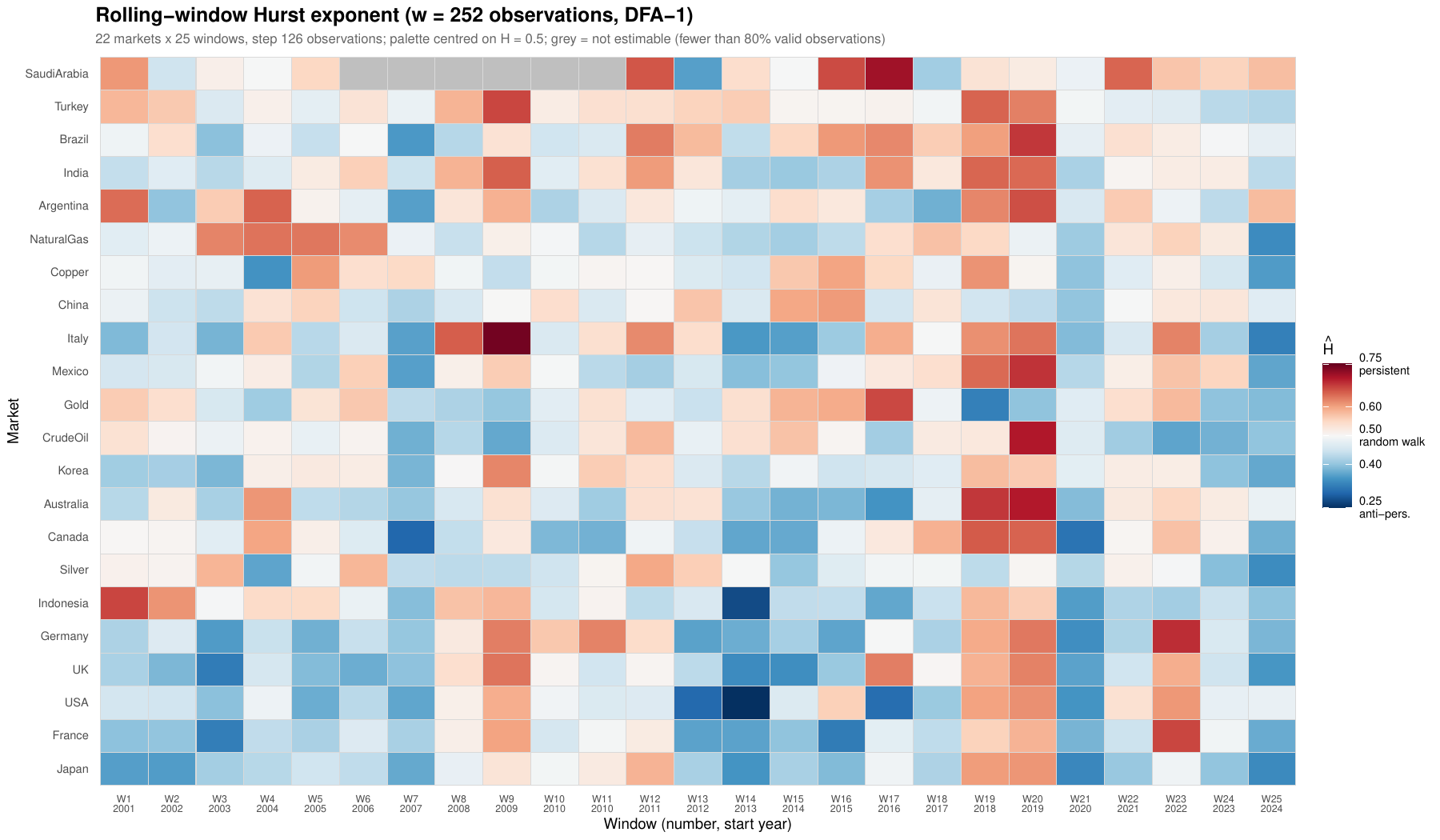}
\caption[Rolling-window Hurst panel]{Rolling-window Hurst exponent
$\hat H_i(t)$ for the \nAssets-asset panel, with $w = \winObs$
observations and DFA-order-1 detrending. The diverging blue-white-red colour scale is
clipped to $[0.25, 0.75]$ and centred on the random-walk benchmark
$H = 0.5$; blue indicates anti-persistence, red indicates persistence.
Grey cells denote insufficient data. The panel-mean $\bar H = \Hbar$
with cross-market standard deviation $\Hsd$ is consistent with weak-
form efficiency in the sense of \citet{fama1970emh}.}
\label{fig:hurst}
\end{figure}

Returns across the \nRowsFmt\ trading dates are leptokurtic in every
series, negatively skewed in \nNegSkew\ of \nAssets, and
Jarque-Bera-rejecting in every series. Per-window ADF tests reject the
unit-root null at conventional levels in every window of the panel
(\nADFreject\ of \nCells\ market-window cells); KPSS tests fail to reject
stationarity in \nKPSSnotreject\ of \nCells. Figure \ref{fig:hurst}
reports the rolling-window Hurst exponent across markets and windows.
The cross-sectional mean of within-market mean $\hat H$ is $\Hbar$ with
a cross-market standard deviation of $\Hsd$; no market sustains
$\hat H \geq 0.55$ across the panel. The most persistent markets on
average are \HtopFive; the most anti-persistent are \HbotFive.
Single-window extremes span \HminCell\ to \HmaxCell.

A cross-section of crisis-period $\hat H$ deltas, defined as the mean
$\hat H$ during the \nCrisisWindowsWord\ crisis windows (GFC \winGFC,
ESDC \winESDC, COVID \winCOVID, Russia-Ukraine \winRU) minus the mean
elsewhere, divides cleanly into two groups. Equity moves toward
persistence in crises (\nEqDeltaPos\ of the \nEqDelta\ equity indices)
-- \deltaPosTop\ -- consistent with trending price dynamics under
crisis-driven flight-to-safety flows. The precious and industrial metals,
natural gas and China move toward anti-persistence -- \deltaNeg\ --
consistent with mean-reverting volatility regimes documented by
\citet{loMacKinlay1988varianceRatio}. The categorical claim that
``crises produce universal anti-persistence'' is not supported by this
panel.

\subsection{HWTE network topology}
\label{subsec:topology}

\begin{figure}[!htbp]
\centering
\includegraphics[width=\textwidth]{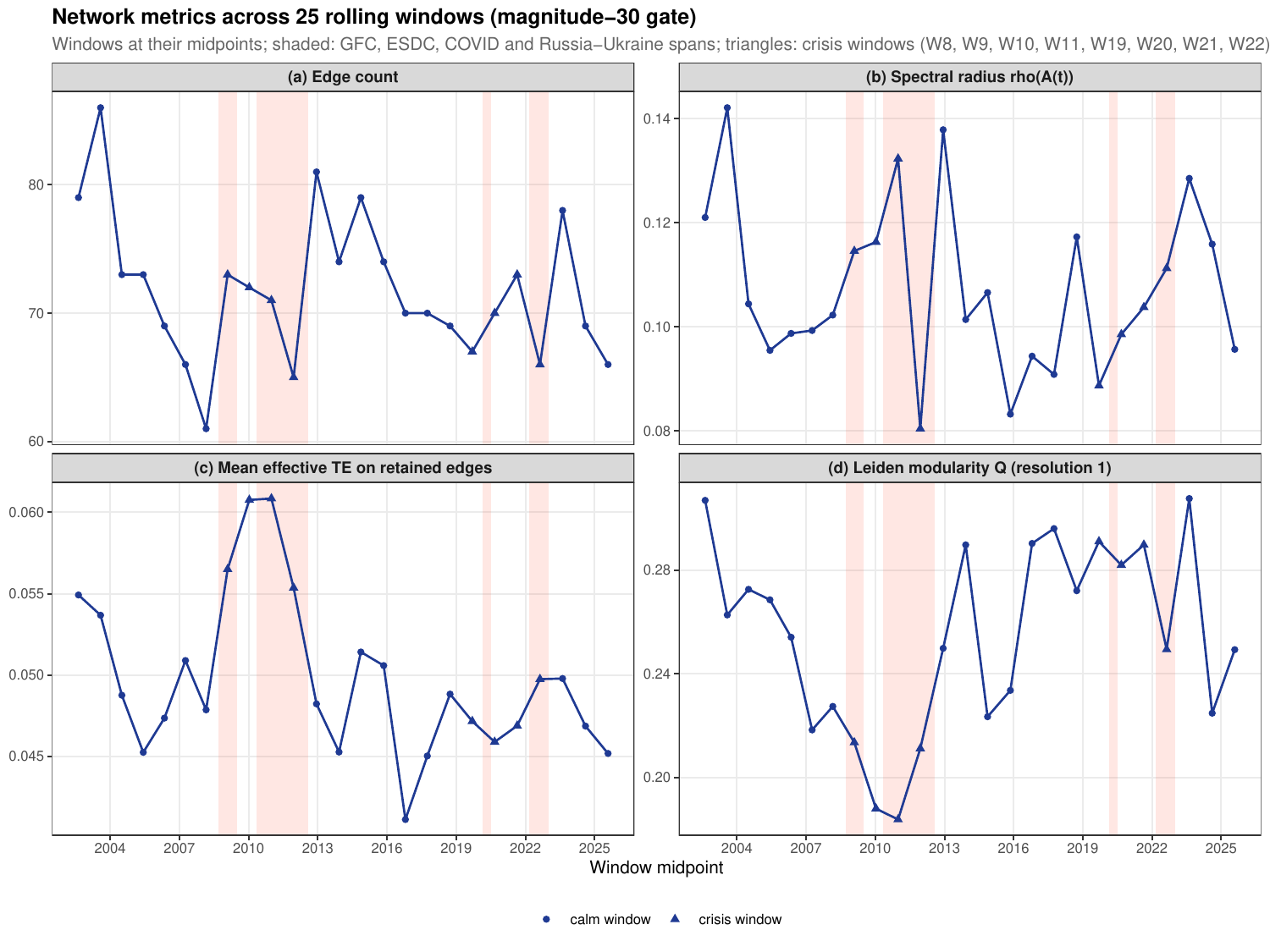}
\caption[HWTE network metrics, magnitude-30 gating]{Time series of HWTE
network metrics under magnitude-30 gating across the \nWindows\ rolling
windows. (a) Retained edge count. (b) Spectral radius $\rho(A(t))$
peaks at $\rhoAmax$ in W\rhoAmaxWindow, a calm window, against a panel
mean of $\rhoAmean$. (c) Mean effective TE on retained edges, peaks at
$\teffPeak$ in \teffPeakWin\ (ESDC), $\teffPeakPct\%$ above the panel
mean of $\teffMean$. (d) Leiden modularity $Q(\gamma{=}1)$, with panel
minimum $\Qmin$ in \QminWin\ (\QminCrisis). Crisis windows \winGFC\
(GFC), \winESDC\ (ESDC), \winCOVID\ (COVID), and \winRU\ (Russia-Ukraine)
are shown with shaded vertical bands.}
\label{fig:metrics}
\end{figure}

Figure \ref{fig:metrics} reports four network metrics across the
\nWindows\ windows under the magnitude-30 gating rule. Network density is
held near-constant at $\densityMean$ (range $\densityMin$-$\densityMax$).
The mean effective TE on retained edges peaks during the ESDC at
$\teffPeak$ nats in \teffPeakDates\ (\teffPeakWin), $\teffPeakPct\%$
above the panel mean of $\teffMean$. The spectral radius $\rho(A(t))$ is
not elevated in the crisis windows (\rhoAcrisisList): its crisis mean of
$\rhoAcrisis$ is indistinguishable from the calm mean of $\rhoAcalm$
(Wilcoxon $p \rhoAcrisisP$), and \nTopThreeRhoCalm\ of its three largest
values fall in calm windows (\rhoAtopThree). The dense-network
amplification phase transition of
\citet{acemogluOzdaglarTahbazSalehi2015systemic} is therefore not
detected as a crisis signature in this panel.

Leiden modularity $Q(\gamma = 1)$ achieves its panel minimum of $\Qmin$
in window \QminWin\ (\QminDates, \QminCrisis) -- well below the panel
mean of $\Qmean$ and the calm-window maximum of $\QcalmMax$
(\QcalmMaxWin, \QcalmMaxDates). The expected ``community structure
collapses in crises'' signature is therefore present in direction but
weak in size: the crisis mean of $\Qcrisis$ lies below the calm mean of
$\Qcalm$ ($\Delta Q = \QdeltaCrisis$, Wilcoxon $p \QcrisisP$). The
$\gamma$-resolution sweep across $\gamma \in \{0.5, 1.0, 1.5, 2.0\}$
shows that the partition is resolution-sensitive: the Adjusted Rand Index
against $\gamma = 1$ exceeds $0.7$ at $\gamma = 1.5$ in \nARIhighOneFive\
of \nWindows\ windows and at $\gamma = 2.0$ in \nARIhighTwo, and
$\gamma = 0.5$ merges the network into at most \maxCommHalf\ communities.
The number of communities at $\gamma = 1$ ranges from \commMin\ to
\commMax\ across the panel (mean $\approx \commMean$).

\subsection{Centrality dynamics: a broadcaster, not an attractor}
\label{subsec:centrality}

\begin{figure}[!htbp]
\centering
\includegraphics[width=\textwidth]{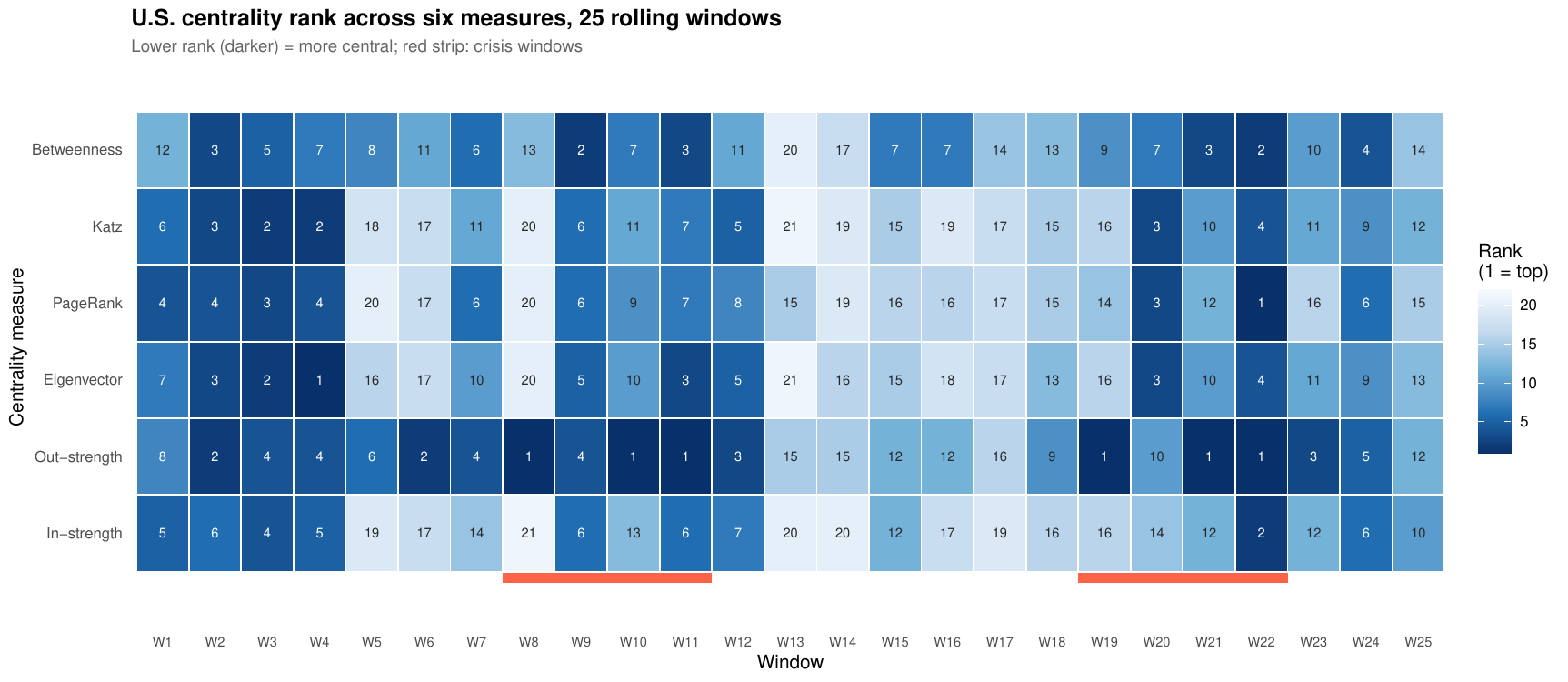}
\caption[U.S.\ centrality rank, 6 measures $\times$ \nWindows\ windows]{U.S.\
rank across six centrality measures (rows) and \nWindows\ windows (columns) on
the magnitude-30 KSG-TE network. Cell colour encodes rank (darker $=$
more central). Rank-1 cells are labelled. The U.S.\ functions as a
\emph{broadcaster} of information in the network and a \emph{cross-
community bridge} during the GFC-ESDC, but not as an information
\emph{attractor} in the in-prestige sense.}
\label{fig:us-ranks}
\end{figure}

Figure \ref{fig:us-ranks} reports the U.S.\ rank across six centrality
measures and \nWindows\ windows. The U.S.\ best eigenvector rank is
\usaBestEigRank\ (\usaBestEigWin, \usaBestEigDates, raw value
$\usaBestEigVal$); during the GFC peak (\gfcPeakWin) the U.S.\ ranks
\usaGfcEigRank\ with eigenvector value $\usaGfcEigVal$. The U.S.\ reaches
rank 1 on eigenvector centrality only in \usaEigOneList\ and on PageRank
only in \usaPrOneList, and never on in-strength (best rank \usaInBest); it
dominates out-strength (rank 1 in \usaOutOneN\ of \nWindows\ windows:
\usaOutOneList, of which \usaOutOneCrisisN\ are crisis windows) and ranks
\usaBetwBest\ on weighted-directed betweenness in \usaBetwBestList.

\begin{figure}[!htbp]
\centering
\includegraphics[width=\textwidth]{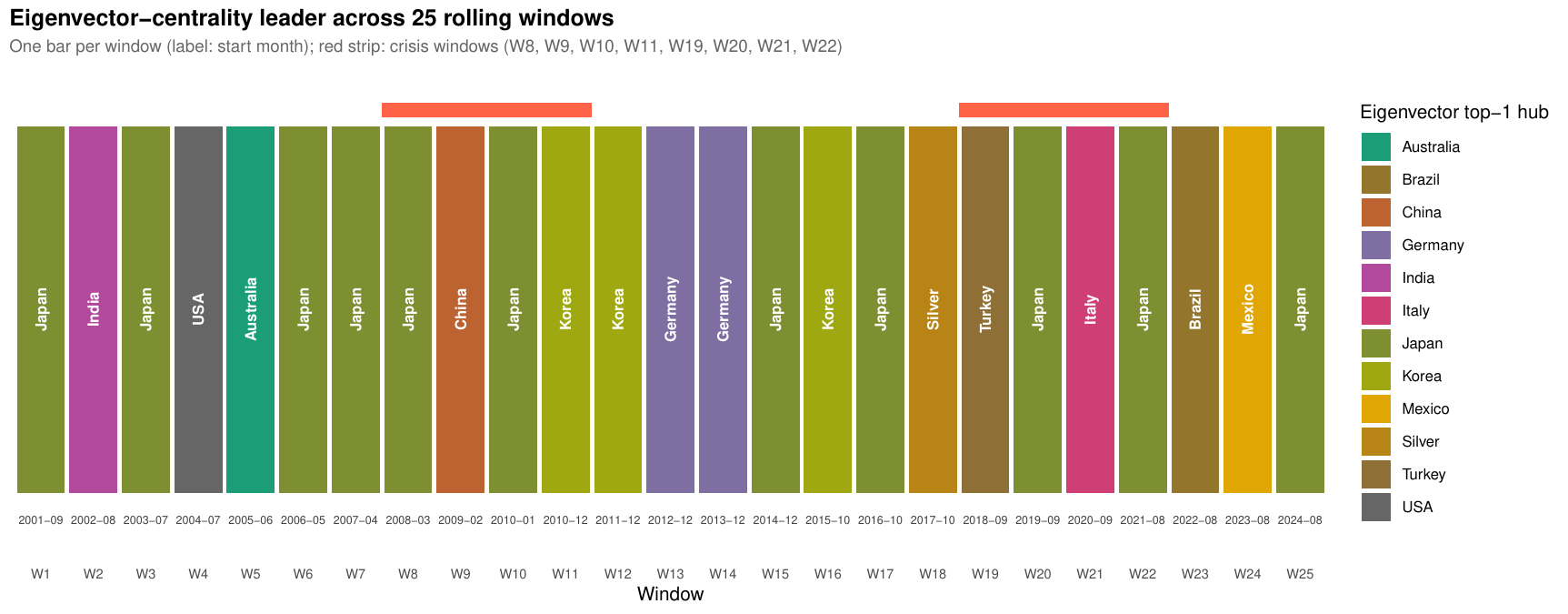}
\caption[Eigenvector hub identity per window]{Identity of the
eigenvector-centrality top-1 market in each rolling window, under
magnitude-30 KSG-TE gating with exponential $\omega$-kernel. Japan
occupies the top position in \hubTopAll\ of \nWindows\ windows and
\hubSecond\ in \hubSecondAll; no other market tops more than two. The
U.S.\ occupies the top position in one window (\usaEigOneList).
Robustness of this finding to the
$\omega$-kernel and to the edge-gating rule is reported in Section
\ref{sec:robust}.}
\label{fig:hubs}
\end{figure}

Figure \ref{fig:hubs} identifies the eigenvector top-1 hub per window.
Across the \nWindows\ windows, Japan occupies the top position in
\hubTopAll\ windows (\hubTopNon\ of the \nNonOverlap\ non-overlapping
windows), and no other market in more than \hubSecondAll: \hubTally. The
same Japan-dominated pattern appears under PageRank and Katz centrality:
Japan tops Katz centrality in \JapanRankOneKatz\ windows and PageRank in
\JapanRankOnePagerank, while the U.S.\ achieves Katz rank 3 or better in
only \usaKatzTopThreeN\ windows and Katz rank $\geq 17$ in \usaKatzLowN. The reading is that the U.S.\ functions as a
\emph{broadcaster} of information and a \emph{cross-community bridge}
during crises, while Japan -- with its anti-correlation to risk-on
flows, deep liquidity, and time-zone alignment with Asian opening
sessions -- attracts the greatest weight of converging information
paths in the eigenvector sense.

\subsection{Network-mediated efficiency}
\label{subsec:network-eff}

\begin{figure}[!htbp]
\centering
\includegraphics[width=\textwidth]{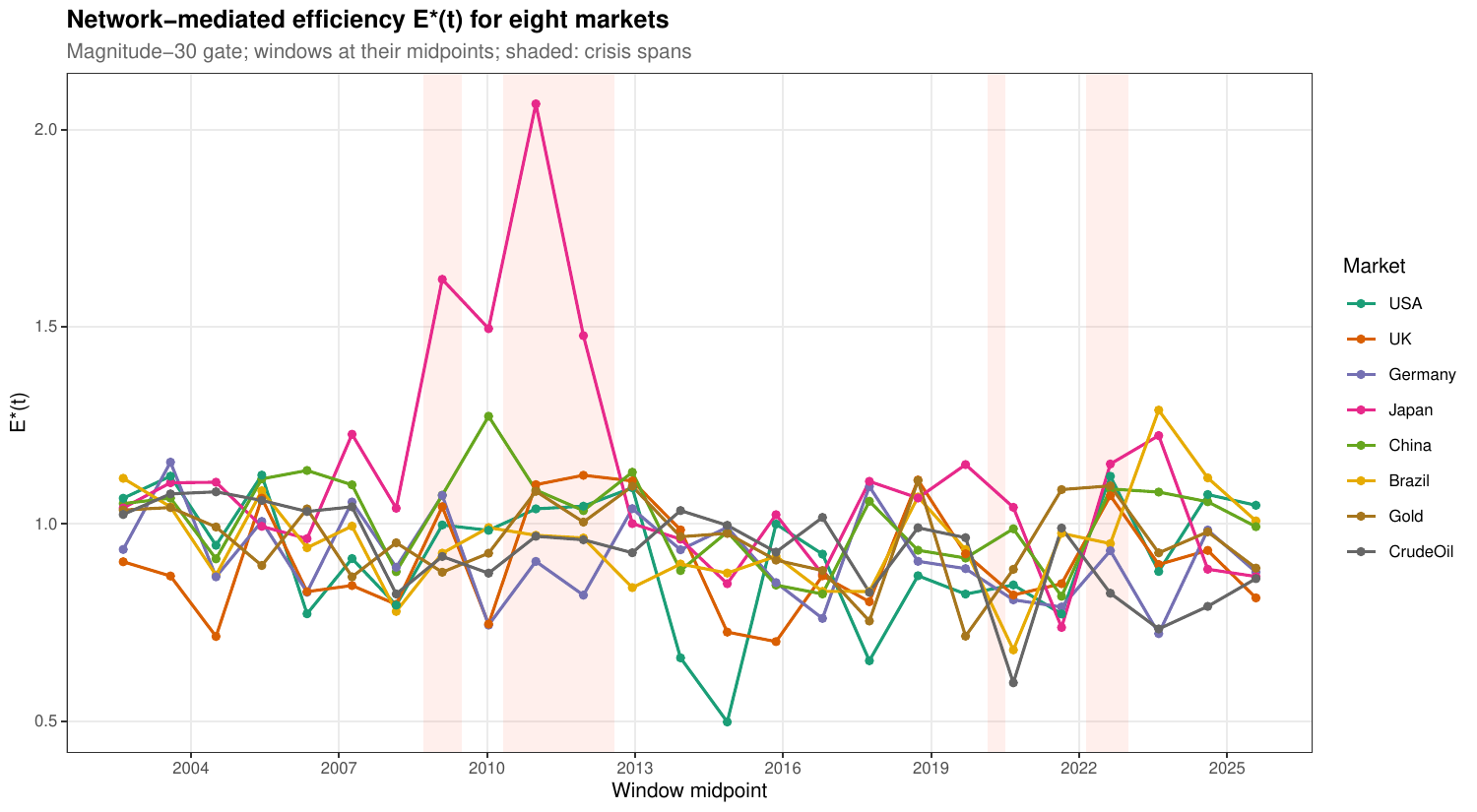}
\caption[Efficiency trajectories]{Network-mediated efficiency
$E_i^*(t)$ trajectories for eight representative markets (USA, UK,
Germany, Japan, China, Brazil, Gold, Crude Oil) under magnitude-30
gating with linear-absolute $\varphi$-kernel. Crisis windows are
shaded.}
\label{fig:efficiency}
\end{figure}

Figure \ref{fig:efficiency} reports the efficiency-vector trajectories
for eight representative markets. The panel-mean $\bar E^*$ averages
$\EstarCalm$ in calm windows, with cross-window dispersion ranging from
$\EstarMin$ (\EstarMinWin, \EstarMinDates) to $\EstarMax$ (\EstarMaxWin,
\EstarMaxDates). Network
feedback amplifies rather than depresses the local efficiency
baseline: the Neumann series $(I - A)^{-1} = \sum_{k \geq 0} A^k$
contributes a strictly positive correction whenever any cross-market
edge is retained. The panel-mean spectral radius of $\rhoAmean$ is well
below the contraction boundary, and the spectral-rescaling safeguard
is never triggered under the magnitude-30 configuration -- the
largest observed $\rho(A(t)) = \rhoAmax$ in W\rhoAmaxWindow\ remains
comfortably inside the Banach contraction ball.

\subsection{Neumann-series convergence and per-asset $\rho$-contribution}
\label{subsec:neumann}

\begin{figure}[!htbp]
\centering
\includegraphics[width=0.72\textwidth]{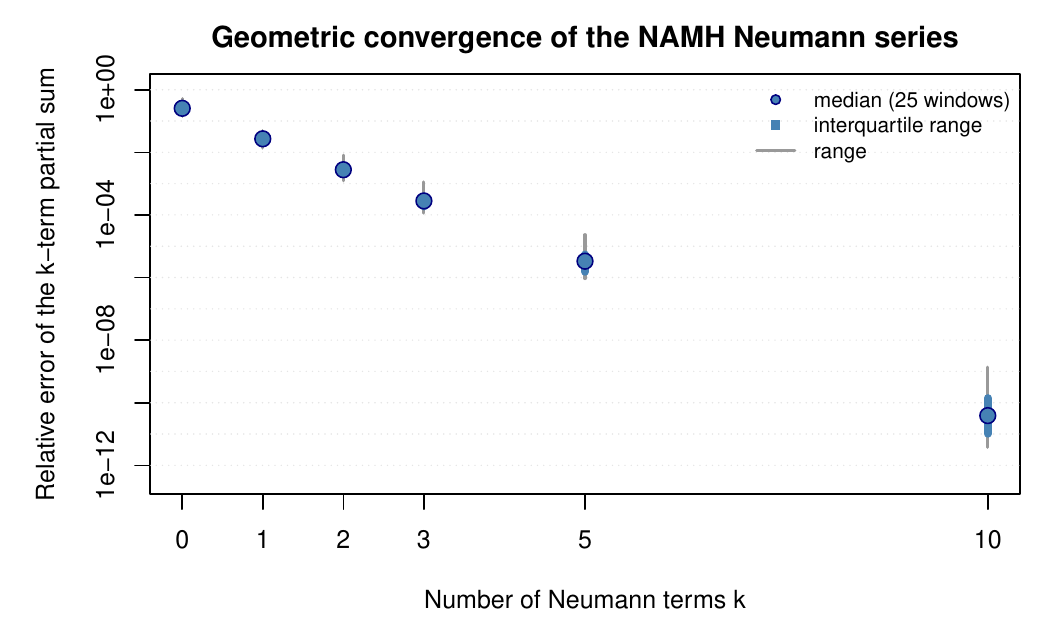}
\caption[Geometric convergence of the Neumann series]{Geometric
convergence of the NAMH Neumann series across the \nWindows\ rolling windows.
Each filled circle is the median across windows of the relative
$\ell^\infty$ error of the partial-sum approximation
$\mathbf{E}^{(k)} = \sum_{j=0}^{k} A^j\,\mathbf{b}$ to the fixed-point
solution. Thick bars give the 25th-75th-percentile range across
windows; thin bars give the min-max range. The geometric decay rate
is consistent with the panel-mean spectral radius $\rho(A) = \rhoAmean$.
Convergence to four-digit accuracy is reached at $k = \neuFourDigitK$
across every window in the panel.}
\label{fig:neumann}
\end{figure}

The fixed-point representation $\mathbf{E}^*(t) = \sum_{k \geq 0}
A(t)^k\,\mathbf{b}(t)$ provides an order-by-order decomposition of
efficiency into the direct local effect ($k = 0$), one-step spillovers
($k = 1$), two-step spillovers ($k = 2$), and higher-order paths.
Figure \ref{fig:neumann} reports the panel-median relative error
$\|\mathbf{E}^{(k)} - \mathbf{E}^*\|_\infty / \|\mathbf{E}^*\|_\infty$
across $k$. Convergence is geometric, as predicted by Theorem
\ref{thm:exist-unique}: the relative error falls from $\neuErrZero$ at
$k = 0$ to $\neuErrOne$ at $k = 1$, \neuErrTwo\ at $k = 2$,
\neuErrThree\ at $k = 3$, and \neuErrTen\ at $k = 10$. The $k = 1$ and
$k = 2$ paths together account for \neuShareTwoPct\% of the
network-amplification term, so the dominant economic content of the
NAMH spillover is two-step cross-market information diffusion.

\begin{figure}[!htbp]
\centering
\includegraphics[width=0.78\textwidth]{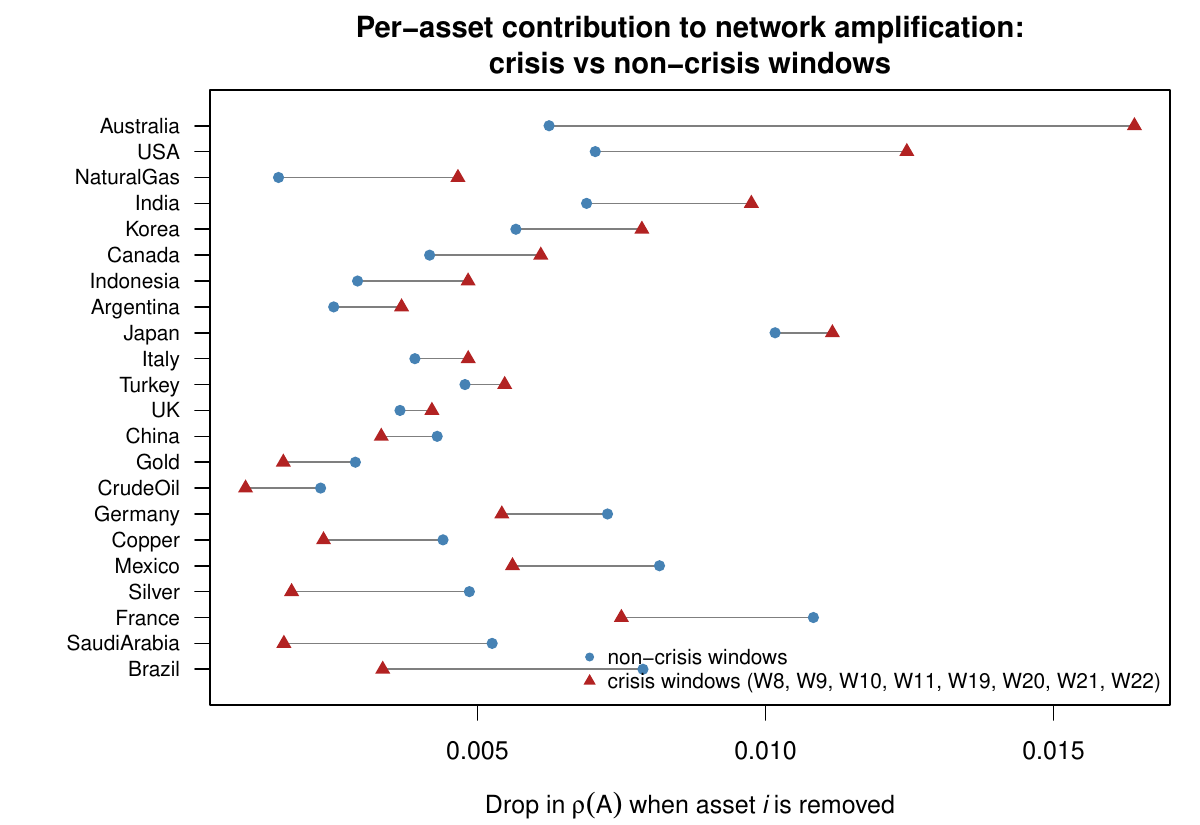}
\caption[Per-asset $\rho(A)$ contribution: crisis vs non-crisis]{
Per-asset contribution to the spectral radius, decomposed by regime.
For each asset $i$ the abscissa is the leave-one-out drop in $\rho(A)$
when row $i$ and column $i$ are removed, averaged separately over
non-crisis windows (blue circles) and crisis windows \winGFC, \winESDC,
\winCOVID, \winRU\ (red triangles). Grey segments connect the two regime means for each
asset. Assets are sorted bottom-to-top by the crisis-minus-non-crisis
difference.}
\label{fig:rho-contrib}
\end{figure}

A complementary decomposition addresses which assets drive the
spectral radius. We compute the leave-one-out spectral-radius drop
$\Delta_i \rho(t) = \rho(A(t)) - \rho(A_{-i}(t))$, where $A_{-i}$
denotes the HWTE matrix with the row and column of asset $i$ deleted,
and report the panel mean of $\Delta_i \rho$ separately for crisis
and non-crisis windows. Figure \ref{fig:rho-contrib} reports the
result. Five markets exhibit crisis-elevated $\rho$-contribution
($\Delta\rho$ in non-crisis $\to$ crisis windows): \rhoUpFive, with
\rhoCrisisTopMkt\ the highest absolute contribution in crises. Because
the spectral radius itself is not higher in crisis windows
(Section~\ref{subsec:topology}), this is a recomposition of the core
rather than an amplification: the crisis-window core is led by
\rhoUpFiveNames. Five markets exhibit the opposite pattern -- their
$\rho$-contribution falls in crisis windows: \rhoDownFive. The crisis
recomposition crowds these contributions toward the periphery of the
network.

\section{Efficiency tiers, connectedness, and the clustered reading}
\label{sec:efficiency-clustering}

Section~\ref{sec:results} characterised network-mediated efficiency
market by market and window by window. This section asks three further
questions. Do markets group into distinct \emph{degrees} of statistical
efficiency? How does the HWTE information network compare with the
variance-decomposition connectedness of
\citet{dieboldYilmaz2012betterUnderstand,dieboldYilmaz2014spillover} on
the same panel, in the classic identifications and in the clustered
identification of \citet{buchwalterDieboldYilmaz2026clustered}? And,
using the algebra of Section~\ref{subsec:theory-clustered}, is the
connectedness network organised by efficiency at all? Every table and
figure below regenerates from the replication materials described in
Appendix~\ref{app:eff-repro}, which first reproduce the canonical
network and fixed point of Section~\ref{sec:results}.

\begin{remark}[Efficiency tiers are not Fama's information taxonomy]
\label{rem:tier-terminology}
``Higher'' and ``lower'' efficiency tiers denote position on the
adaptive-markets continuum of statistical (long-memory) efficiency:
distance of $H$ from $0.5$ and of $\mathbf{E}^*$ from its own upper range.
They are not the weak-, semi-strong- and strong-form classes of
\citet{fama1970emh}, which classify the information set priced into an
asset rather than the persistence of its return process. We use the
labels Tier~I to Tier~V to keep the two apart.
\end{remark}

\subsection{Data-driven efficiency tiers}
\label{subsec:eff-tiers}

For each market of the \nAssets-asset panel (Saudi Arabia, absent
from windows \saudiAbsentWindows, enters with the windows in which its
Hurst exponent is estimable; Section~\ref{subsec:eff-robust} examines
the choice) we form the feature vector $(\bar H_i,\ \mathrm{sd}(H_i),\
\bar E^*_i,\ \mathrm{sd}(E^*_i))$ across the \nWindows\ windows of
Figures~\ref{fig:hurst} and~\ref{fig:efficiency}, and cluster with three
methods: a Gaussian mixture with model and $K$ chosen by BIC over
$K\in\{2,\ldots,5\}$ \citep{fraleyRaftery2002mclust}, Ward's hierarchical
method on standardised features \citep{ward1963hierarchical}, and
$K$-means with 50 restarts. BIC selects $K = \bicK$; a market-by-market
majority vote across the three methods populates all \nTiersWord\
candidate tiers, labelled by ascending mean $|\bar H_i - 0.5|$:
Tier~I ($N=\nTierI$), Tier~II ($N=\nTierII$), Tier~III ($N=\nTierIII$),
Tier~IV ($N=\nTierIV$) and Tier~V ($N=\nTierV$, Japan alone).
Method agreement is imperfect and is reported rather than smoothed
(Table~\ref{tab:v4-ari}): the Adjusted Rand Index
\citep{hubertArabie1985comparingPartitions} is \ariGmmHier\ between the
mixture and Ward, \ariGmmKm\ between the mixture and $K$-means, and
\ariHierKm\ between Ward and $K$-means. The majority-vote consensus, not
any single method, is the headline partition.

\begin{table}[htbp]
\centering
\caption{Adjusted Rand Index across the three clustering methods.}
\label{tab:v4-ari}
\begin{tabular}{lccc}
\toprule
 & GMM & Hierarchical & $K$-means \\
\midrule
GMM & 1.000 & 0.415 & 0.415 \\
Hierarchical & 0.415 & 1.000 & 1.000 \\
$K$-means & 0.415 & 1.000 & 1.000 \\
\bottomrule
\end{tabular}
\end{table}

Table~\ref{tab:v4-eff-tiers} lists the tiers with each market's
rolling-window Hurst mean and dispersion, mean network-mediated
efficiency, and a full-sample test of $H_0\!:\!H_i = 0.5$ that locates
the full-sample estimate in its fixed-length moving block bootstrap
distribution under a short-memory null \citep{kunsch1989blockBootstrap}
($B = \hBootB$, block length \hBootBlock) with Benjamini--Hochberg control
across the \nMarketsBoot\ tests. \nRejectNull\ of \nMarketsBoot\ markets
reject at $q < 0.05$, and \nRawNull\ at the nominal $p < 0.05$. The
null result and the tiering are not in tension: no single market's
\emph{static, full-sample} exponent is distinguishable from $0.5$, which
is the motivation for the rolling, network-mediated construction of
Section~\ref{sec:framework} in the first place. The tiers rank markets by
their regime-local behaviour, not by a static deviation.

\begin{table}[htbp]
\centering
\small
\caption{Efficiency tiers (majority-vote consensus). $\bar H_i$,
$\mathrm{sd}(H_i)$ and $\bar E^*_i$ are across the \nWindows\ windows;
$\hat H_i^{\mathrm{full}}$ is the full-sample DFA-1 estimate, null centre
the mean of its moving-block-bootstrap distribution under the
short-memory null, $p$ the two-sided bootstrap $p$-value for
$H_0\!:\!H_i=0.5$, and $q$ its Benjamini--Hochberg adjustment.}
\label{tab:v4-eff-tiers}
\begin{tabular}{llccccccc}
\toprule
Market & Tier & $\bar H_i$ & $\mathrm{sd}(H_i)$ & $\bar E^*_i$ & $\hat H_i^{\mathrm{full}}$ & null centre & $p$ & BH $q$ \\
\midrule
Mexico & I & 0.491 & 0.080 & 0.993 & 0.497 & 0.501 & 0.912 & 0.970 \\
India & I & 0.512 & 0.078 & 1.005 & 0.522 & 0.512 & 0.723 & 0.970 \\
Korea & I & 0.483 & 0.066 & 1.069 & 0.510 & 0.512 & 0.970 & 0.970 \\
Australia & I & 0.482 & 0.090 & 1.024 & 0.502 & 0.483 & 0.530 & 0.970 \\
\midrule
Copper & II & 0.494 & 0.067 & 0.964 & 0.537 & 0.531 & 0.876 & 0.970 \\
China & II & 0.493 & 0.055 & 1.013 & 0.589 & 0.542 & 0.151 & 0.970 \\
Silver & II & 0.477 & 0.066 & 0.983 & 0.457 & 0.474 & 0.600 & 0.970 \\
Turkey & II & 0.527 & 0.062 & 0.987 & 0.496 & 0.499 & 0.911 & 0.970 \\
\midrule
Natural Gas & III & 0.502 & 0.077 & 0.923 & 0.455 & 0.466 & 0.700 & 0.970 \\
Gold & III & 0.490 & 0.082 & 0.960 & 0.447 & 0.462 & 0.588 & 0.970 \\
Argentina & III & 0.511 & 0.084 & 0.931 & 0.501 & 0.499 & 0.933 & 0.970 \\
Brazil & III & 0.515 & 0.078 & 0.956 & 0.500 & 0.499 & 0.947 & 0.970 \\
Crude Oil & III & 0.483 & 0.078 & 0.934 & 0.470 & 0.493 & 0.586 & 0.970 \\
Saudi Arabia & III & 0.545 & 0.088 & 0.932 & 0.567 & 0.537 & 0.410 & 0.970 \\
\midrule
Italy & IV & 0.492 & 0.114 & 0.884 & 0.496 & 0.483 & 0.659 & 0.970 \\
Canada & IV & 0.478 & 0.095 & 0.935 & 0.498 & 0.491 & 0.832 & 0.970 \\
Indonesia & IV & 0.476 & 0.091 & 0.958 & 0.534 & 0.543 & 0.807 & 0.970 \\
Germany & IV & 0.471 & 0.101 & 0.914 & 0.509 & 0.482 & 0.402 & 0.970 \\
UK & IV & 0.459 & 0.097 & 0.906 & 0.466 & 0.444 & 0.447 & 0.970 \\
USA & IV & 0.458 & 0.099 & 0.923 & 0.497 & 0.476 & 0.539 & 0.970 \\
France & IV & 0.453 & 0.086 & 0.955 & 0.489 & 0.464 & 0.419 & 0.970 \\
\midrule
Japan & V & 0.445 & 0.080 & 1.123 & 0.492 & 0.487 & 0.894 & 0.970 \\
\bottomrule
\end{tabular}
\end{table}

\subsection{Variance-decomposition connectedness on the NAMH panel}
\label{subsec:classic-dy}

Common trading dates one to several sessions apart make the daily
returns an irregularly spaced series, unsuited to a joint VAR on
\nAssets\ markets. We cumulate to weekly log returns -- the convention of
\citet{buchwalterDieboldYilmaz2026clustered} -- which leaves
$T = \nWeeksFmt$ complete weeks over 2001--2026 and leaves the daily
Hurst/HWTE machinery of Sections~\ref{sec:framework}--\ref{sec:results}
untouched. AIC selects a VAR(\varLag); decompositions use the $H = 10$-week
horizon with $H = 12$ as a sensitivity check. Table~\ref{tab:v4-dy}
reports the generalised decomposition
\citep{koopPesaranPotter1996,pesaranShin1998}. Total connectedness is
\Cgen\ under the generalised identification and \Corth\ under a single
Cholesky ordering, the two polar cases of Proposition~\ref{prop:block-orth}.
The largest net transmitters are \topNetTransmitters; the largest net
receivers are \topNetReceivers\ -- a core-periphery pattern consistent
with the broadcaster role of the USA in Section~\ref{subsec:centrality}.

\begin{table}[htbp]
\centering
\small
\caption{Generalised connectedness, full sample (weekly, $N = \nAssets$,
VAR(\varLag), $H = 10$). FROM: share of own forecast-error variance due to
others; TO: share of others' variance due to this market; NET $=$ TO $-$
FROM. Sorted by NET.}
\label{tab:v4-dy}
\begin{tabular}{lrrr}
\toprule
Market & FROM & TO & NET \\
\midrule
Canada & 86.4 & 126.1 & $+39.7$ \\
France & 87.2 & 122.7 & $+35.5$ \\
UK & 86.5 & 116.7 & $+30.2$ \\
USA & 86.1 & 115.5 & $+29.4$ \\
Germany & 86.6 & 115.8 & $+29.2$ \\
Italy & 85.8 & 108.4 & $+22.6$ \\
Australia & 83.8 & 95.5 & $+11.7$ \\
Brazil & 82.1 & 93.0 & $+10.9$ \\
Mexico & 82.9 & 91.8 & $+8.9$ \\
Korea & 81.7 & 83.3 & $+1.5$ \\
Japan & 81.1 & 79.8 & $-1.2$ \\
India & 79.5 & 71.7 & $-7.8$ \\
Silver & 64.9 & 56.6 & $-8.3$ \\
Natural Gas & 12.8 & 4.1 & $-8.7$ \\
Indonesia & 76.1 & 62.7 & $-13.5$ \\
Copper & 69.0 & 51.7 & $-17.3$ \\
Gold & 51.0 & 31.8 & $-19.3$ \\
Argentina & 69.7 & 44.1 & $-25.6$ \\
Crude Oil & 48.6 & 22.1 & $-26.4$ \\
Turkey & 68.4 & 39.4 & $-29.0$ \\
China & 48.3 & 18.7 & $-29.6$ \\
Saudi Arabia & 50.5 & 17.8 & $-32.7$ \\
\bottomrule
\end{tabular}
\end{table}

\subsection{Clustered connectedness with efficiency-tier blocks}
\label{subsec:clustered-dy}

With the efficiency tiers as the block partition, the block-orthogonalised
identification of Proposition~\ref{prop:block-orth}, averaged over the
$\nTiers! = \nOrderings$ tier orderings, gives total connectedness \Cclust\ -- between the
generalised value \Cgen\ and the orthogonalised value \Corth, the
bracketing that \citet{buchwalterDieboldYilmaz2026clustered} report for
their regional clusters. Table~\ref{tab:v4-bootstrap} attaches sampling
uncertainty by a moving-block residual bootstrap of the VAR
($B = \bootB$, block length four weeks, which preserves the conditional
heteroskedasticity of weekly residuals): the three indices are separated
by several bootstrap standard deviations and the ordering
$\mathcal{C}^{\mathrm{orth}} \leq \mathcal{C}^{\mathrm{clust}} \leq
\mathcal{C}^{\mathrm{gen}}$ holds in a fraction \bootBracketHten\ of
bootstrap draws at $H = 10$ (\bootBracketHtwelve\ at $H = 12$). Consistent
with Remark~\ref{rem:no-bracketing-theorem}, this is an empirical
regularity of the panel rather than a property of the construction. The
horizon is immaterial: VAR(1) coefficients on weekly returns are small
enough that the decompositions have converged well before ten weeks, and
the $H = 12$ values coincide with the $H = 10$ values to the reported
precision.

\begin{table}[htbp]
\centering
\caption{Total connectedness under three identifications with
moving-block residual-bootstrap intervals ($B = \bootB$).}
\label{tab:v4-bootstrap}
\begin{tabular}{clccc}
\toprule
$H$ & Identification & $\mathcal{C}$ & boot.\ s.d. & 90\% CI \\
\midrule
10 & generalised ($C=1$) & 71.3 & 2.15 & [67.8, 74.9] \\
10 & efficiency-tier clustered ($C=5$) & 54.4 & 2.24 & [51.7, 59.1] \\
10 & orthogonalised ($C=N$) & 43.7 & 2.11 & [41.9, 48.7] \\
\midrule
12 & generalised ($C=1$) & 71.3 & 2.15 & [67.8, 74.9] \\
12 & efficiency-tier clustered ($C=5$) & 54.4 & 2.24 & [51.7, 59.1] \\
12 & orthogonalised ($C=N$) & 43.7 & 2.11 & [41.9, 48.7] \\
\bottomrule
\end{tabular}
\end{table}

\begin{figure}[htbp]
\centering
\includegraphics[width=\textwidth]{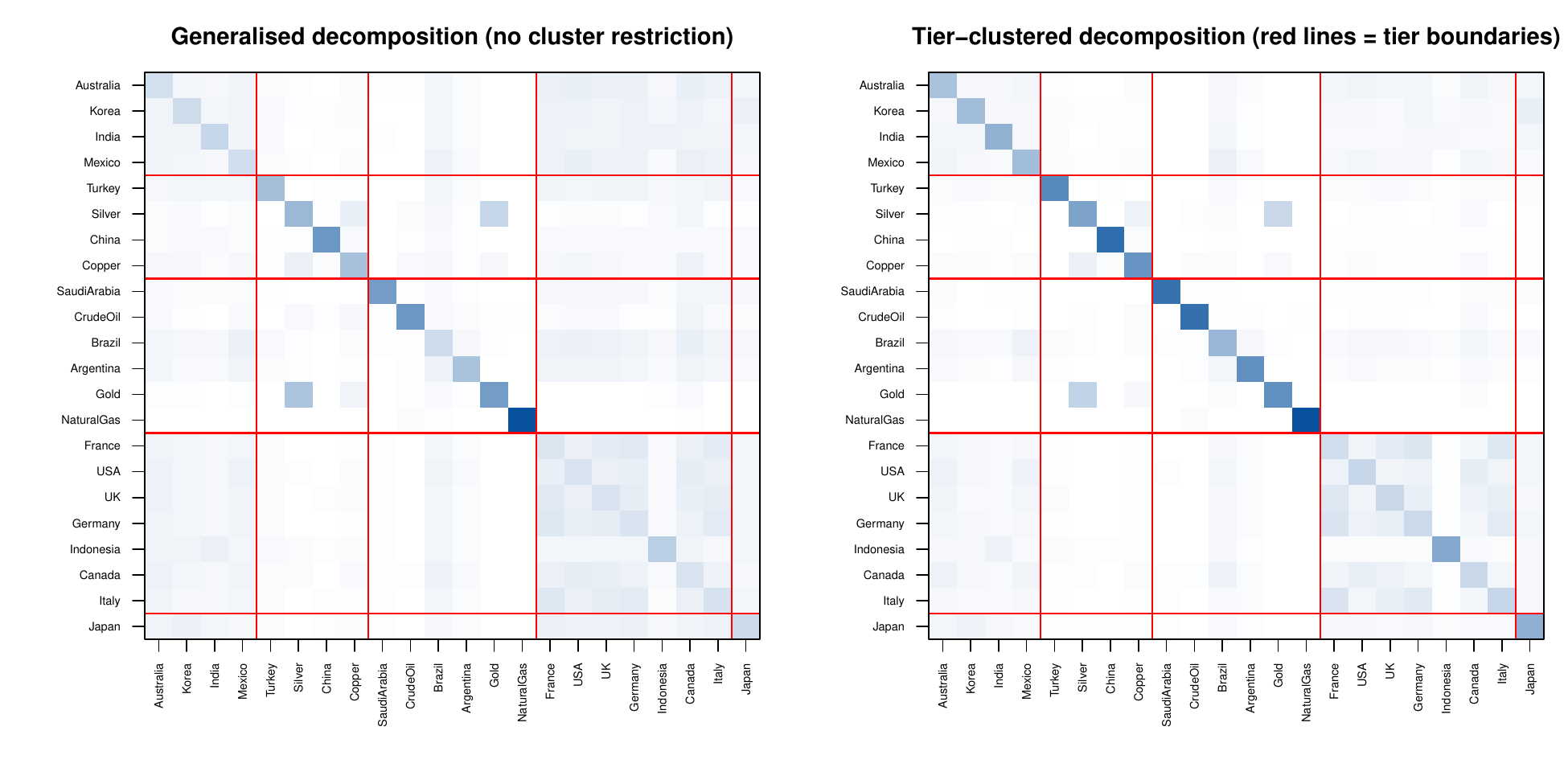}
\caption{Pairwise connectedness $\theta_{ij}(H)$, generalised (left) and
efficiency-tier clustered (right), markets ordered by tier; lines mark tier
boundaries.}
\label{fig:heatmap-block}
\end{figure}

Table~\ref{tab:v4-occ} gives the own / co-movement / contagion
decomposition of Definition~\ref{def:share-matrix} by tier, with
bootstrap intervals; the identity of Proposition~\ref{prop:partition-unity}
holds to numerical precision in every row. Tier~III, of intermediate
efficiency, carries the highest own share (\ownIII) and the lowest
within-tier co-movement among the multi-market tiers (\comoveIII);
Tier~IV, the large developed equity markets, carries the lowest own share
(\ownIV) and the highest co-movement (\comoveIV). Tier~V, the least
efficient, is Japan alone, so its within-tier co-movement is zero by
construction; its own share (\ownV) exceeds that of Tier~I, the most
efficient (\ownI), in a fraction \bootOwnLastGtFirst\ of bootstrap draws.
Read plainly: idiosyncrasy is not monotone in statistical efficiency, and
the least efficient multi-market tier moves together with same-tier peers
the most -- the reverse of a naive prior that markets of similar
efficiency should be the most ``independent''. Figure~\ref{fig:heatmap-block}
shows the block structure against the non-clustered decomposition.
Identification also reshapes individual roles: \swingMkt's net position
moves from \swingNetGen\ under the generalised identification to
\swingNetClust\ under the tier-clustered one, the largest swing in the
panel and an echo, from a different clustering criterion, of the
identification-dependence of hub status that
\citet{buchwalterDieboldYilmaz2026clustered} emphasise for the USA.

\begin{table}[htbp]
\centering
\small
\caption{Own / co-movement / contagion decomposition by efficiency tier
(clustered identification, full sample) with 90\% moving-block
residual-bootstrap intervals.}
\label{tab:v4-occ}
\begin{tabular}{lcccc}
\toprule
Tier & $N_c$ & $\Theta_c^{own}$ [90\% CI] & $\Theta_c^{comove}$ [90\% CI] & $\Theta_{c\leftarrow\bullet}^{contag}$ [90\% CI] \\
\midrule
I (most efficient) & 4 & 0.361 [0.316, 0.399] & 0.139 [0.126, 0.152] & 0.500 [0.466, 0.541] \\
II & 4 & 0.616 [0.560, 0.643] & 0.064 [0.056, 0.077] & 0.320 [0.295, 0.372] \\
III & 6 & 0.667 [0.600, 0.695] & 0.056 [0.050, 0.076] & 0.277 [0.252, 0.328] \\
IV & 7 & 0.242 [0.212, 0.268] & 0.395 [0.375, 0.410] & 0.363 [0.331, 0.402] \\
V (least efficient) & 1 & 0.426 [0.373, 0.467] & 0.000 [0.000, 0.000] & 0.574 [0.533, 0.627] \\
\midrule
System & 22 & 0.462 [0.415, 0.493] & 0.131 [0.127, 0.137] & 0.407 [0.377, 0.451] \\
\bottomrule
\end{tabular}
\end{table}

\subsection{The NAMH tier decomposition}
\label{subsec:namh-tier-results}

\begin{figure}[htbp]
\centering
\includegraphics[width=\textwidth]{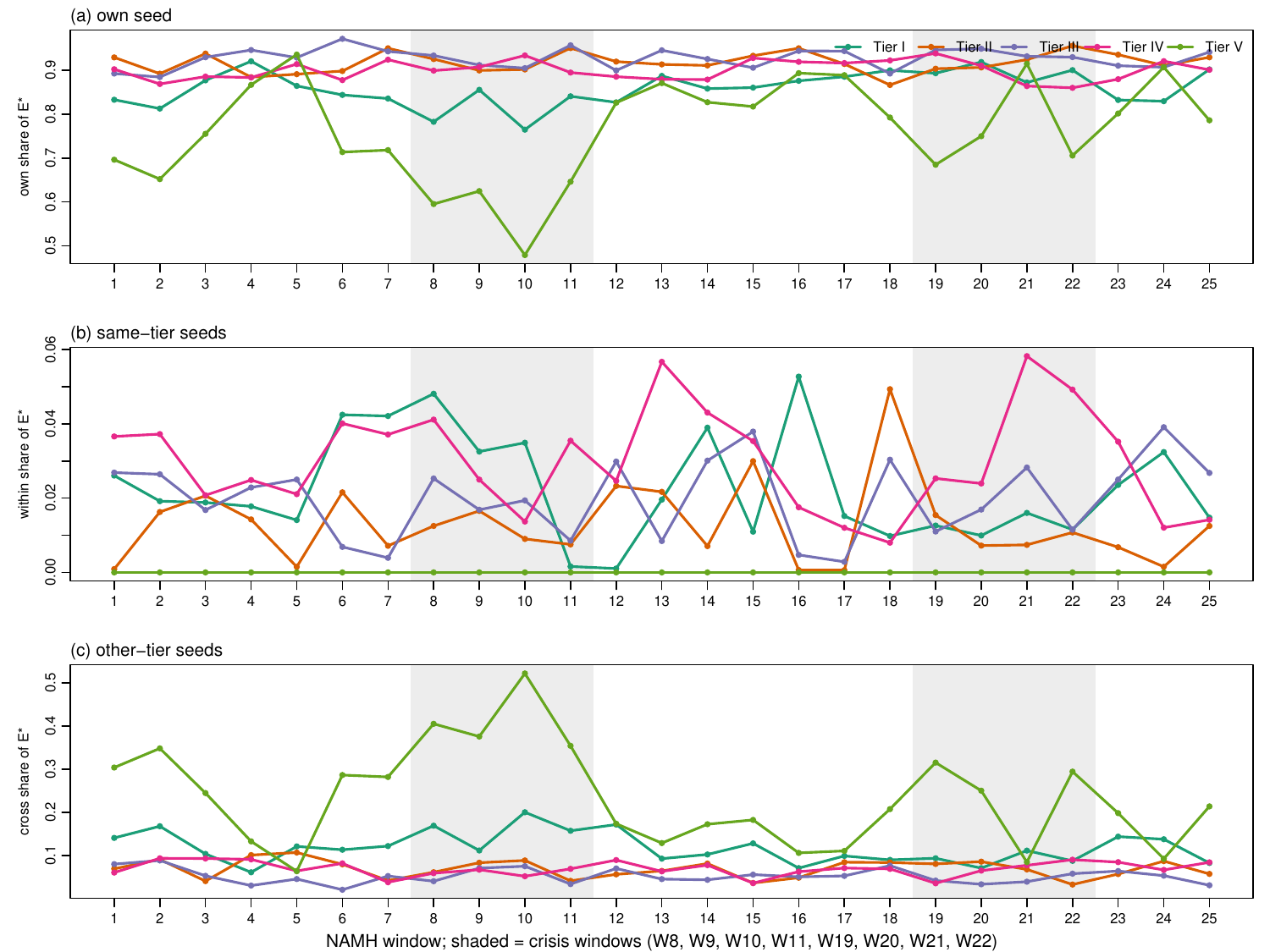}
\caption{NAMH tier decomposition across the \nWindows\ windows: tier means
of the own (a), same-tier (b) and other-tier (c) shares of $E^*_i$;
shaded windows are the crisis anchors.}
\label{fig:namh-tier-decomp}
\end{figure}

\begin{figure}[htbp]
\centering
\includegraphics[width=0.95\textwidth]{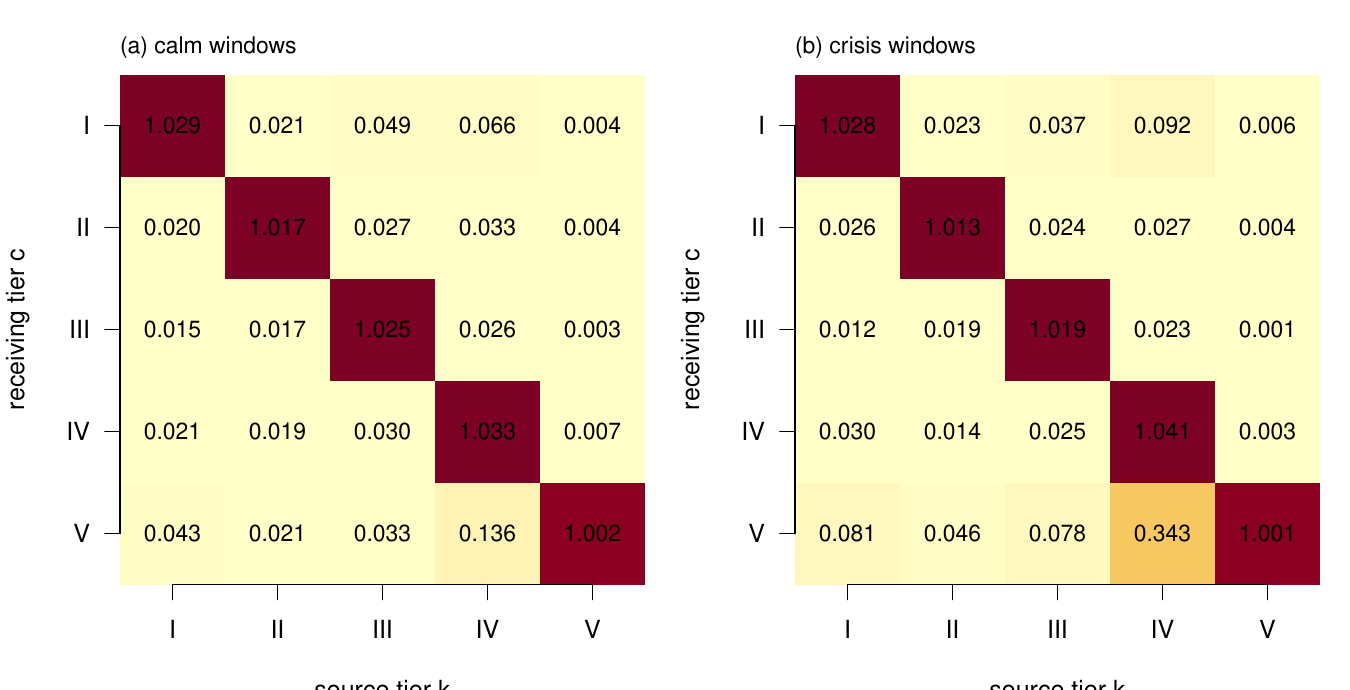}
\caption{Tier multiplier matrices $\Psi$ in calm (a) and crisis (b)
windows.}
\label{fig:psi}
\end{figure}

Proposition~\ref{prop:neumann-tiers} gives the efficiency fixed point the
same accounting. For every window we recompute $A(t)$ and
$\mathbf{E}^*(t)$ from primitives (they match the values behind
Figure~\ref{fig:efficiency} to \decompMaxDE) and split each market's
$E^*_i$ into the share generated by its own seed $\varphi(H_i)$, by
same-tier seeds, and by other-tier seeds. Table~\ref{tab:v4-namh-tiers}
reports tier means by regime (crisis windows \winGFC, \winESDC,
\winCOVID\ and \winRU\ as in Section~\ref{subsec:neumann}) and
Figure~\ref{fig:namh-tier-decomp} the full window paths. Three features
stand out. First, the own share lies between $\namhOwnMin$ and
$\namhOwnMax$ in every tier and regime: network amplification adds
$\namhAmpMin$--$\namhAmpMax\%$ to local efficiency, the order of magnitude
already implied by the spectral radius of Section~\ref{subsec:network-eff}.
Second, the cross-section agrees with the variance decomposition:
Tier~\idioTierNAMH\ has the highest own share and the smallest network
component, so the markets of intermediate efficiency are the most
idiosyncratic under \emph{both} accountings, while Tier~\lowOwnTierNAMH,
Japan, has the lowest own share. Third, the regime contrast is
tier-specific. The other-tier share of Tier~V rises from \crossCalmAllV\
in calm windows to \crossCrisisAllV\ in crisis windows (Welch
$p \crossPAllV$), that of Tier~I from \crossCalmAllI\ to \crossCrisisAllI\
($p \crossPAllI$), while Tiers~II to~IV are flat or fall slightly
(Tier~IV \crossCalmAllIV\ to \crossCrisisAllIV, $p \crossPAllIV$). With
\nCrisisWindowsWord\ crisis windows these contrasts are indicative rather
than decisive, and we report them as such. Table~\ref{tab:v4-psi} gives
the tier multiplier matrices $\Psi$ of
Proposition~\ref{prop:neumann-tiers}(iii): the diagonal
\psiDiagAllAboveOne, the off-diagonal multipliers into \psiRiseTiers\
rise in crises -- most strongly from Tier~IV into Tier~V, from
\psiVIVcalm\ to \psiVIVcrisis\ -- and those into the other tiers do not
(Figure~\ref{fig:psi}).

\begin{table}[htbp]
\centering
\small
\caption{NAMH tier decomposition of $\mathbf{E}^*$: tier means of the
own, same-tier and other-tier shares and of the amplification
$\bar E^*/\bar\varphi - 1$, by regime; last column: crisis minus calm
other-tier share with Welch $p$-value.}
\label{tab:v4-namh-tiers}
\begin{tabular}{llcccccc}
\toprule
Tier & Regime & Windows & $s^{own}$ & $s^{within}$ & $s^{cross}$ & $\bar E^*/\bar\varphi - 1$ & $\Delta s^{cross}$ ($p$) \\
\midrule
I & calm & 17 & 0.862 & 0.023 & 0.115 & 0.168 &  \\
 & crisis & 8 & 0.854 & 0.021 & 0.125 & 0.187 & $+0.011$ (0.564) \\
\midrule
II & calm & 17 & 0.916 & 0.014 & 0.070 & 0.097 &  \\
 & crisis & 8 & 0.921 & 0.011 & 0.068 & 0.091 & $-0.002$ (0.818) \\
\midrule
III & calm & 17 & 0.925 & 0.021 & 0.054 & 0.086 &  \\
 & crisis & 8 & 0.934 & 0.017 & 0.049 & 0.074 & $-0.005$ (0.535) \\
\midrule
IV & calm & 17 & 0.900 & 0.028 & 0.072 & 0.116 &  \\
 & crisis & 8 & 0.901 & 0.034 & 0.065 & 0.116 & $-0.008$ (0.299) \\
\midrule
V & calm & 17 & 0.809 & 0.000 & 0.191 & 0.251 &  \\
 & crisis & 8 & 0.675 & 0.000 & 0.325 & 0.529 & $+0.134$ (0.021) \\
\bottomrule
\end{tabular}
\end{table}

\begin{table}[htbp]
\centering
\caption{Tier multiplier matrices $\Psi_{ck} = N_c^{-1}\sum_{i\in c}\sum_{j\in k}M_{ij}$,
averaged over calm windows (left block) and crisis windows (right block).}
\label{tab:v4-psi}
\begin{tabular}{lcccccccccc}
\toprule
Receiving tier $c$ & $k=$I & $k=$II & $k=$III & $k=$IV & $k=$V & $k=$I & $k=$II & $k=$III & $k=$IV & $k=$V \\
\midrule
I & 1.029 & 0.021 & 0.049 & 0.066 & 0.004 & 1.028 & 0.023 & 0.037 & 0.092 & 0.006 \\
II & 0.020 & 1.017 & 0.027 & 0.033 & 0.004 & 0.026 & 1.013 & 0.024 & 0.027 & 0.004 \\
III & 0.015 & 0.017 & 1.025 & 0.026 & 0.003 & 0.012 & 0.019 & 1.019 & 0.023 & 0.001 \\
IV & 0.021 & 0.019 & 0.030 & 1.033 & 0.007 & 0.030 & 0.014 & 0.025 & 1.041 & 0.003 \\
V & 0.043 & 0.021 & 0.033 & 0.136 & 1.002 & 0.081 & 0.046 & 0.078 & 0.343 & 1.001 \\
\bottomrule
\end{tabular}
\end{table}

\subsection{Rolling dynamics and the crisis signature}
\label{subsec:rolling}

\begin{figure}[htbp]
\centering
\includegraphics[width=\textwidth]{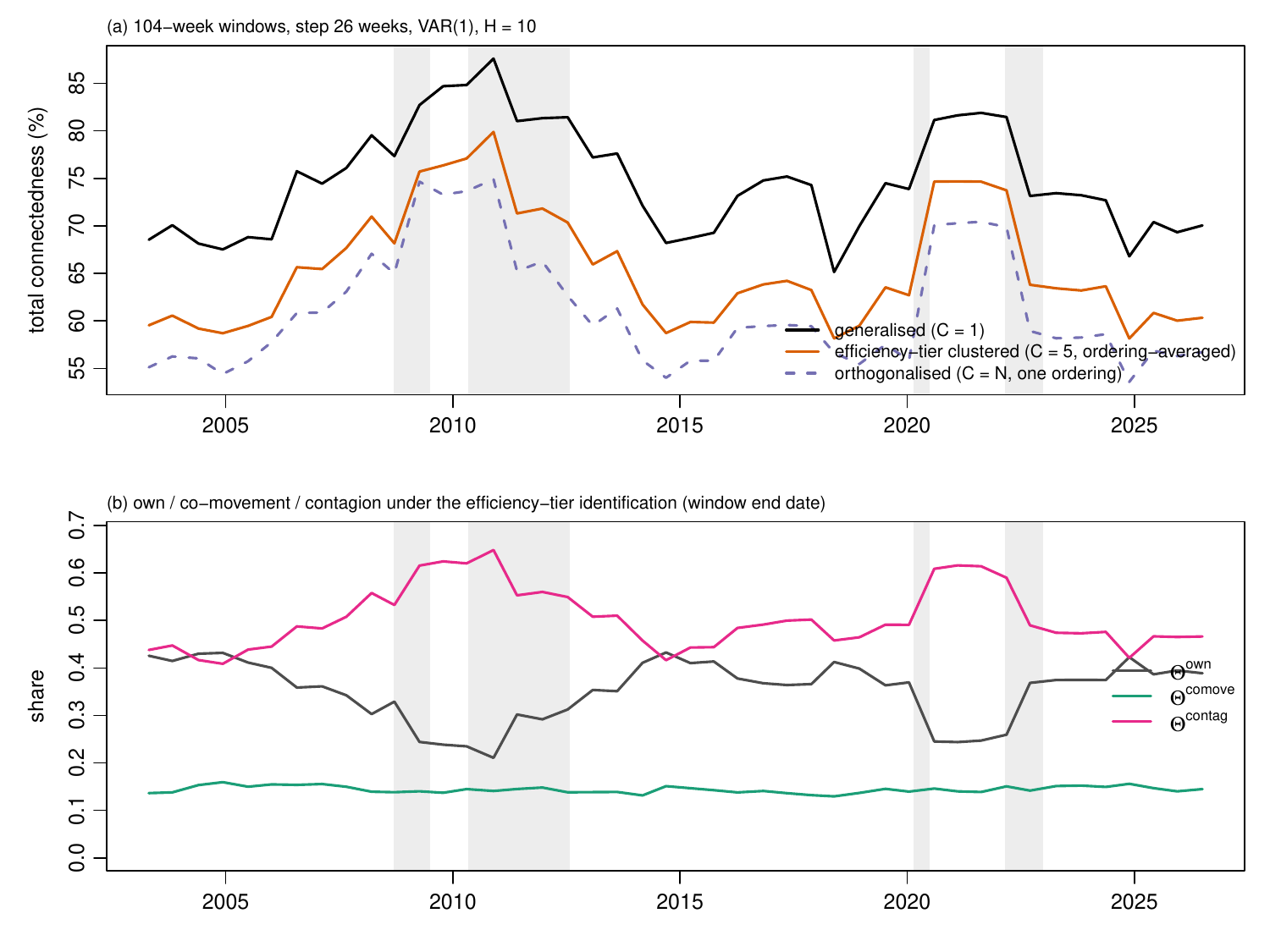}
\caption{Rolling connectedness (104-week windows, quarterly step): (a) the
three identifications; (b) own, co-movement and contagion shares under
the efficiency-tier identification. Shaded spans: GFC, European
sovereign-debt crisis, COVID-19, Russia--Ukraine.}
\label{fig:rolling}
\end{figure}

Figure~\ref{fig:rolling} tracks the three identifications on 104-week
windows stepped quarterly (\nRollWindows\ windows, VAR(1) throughout;
Appendix~\ref{app:eff-repro} explains why the order is held fixed). The
bracketing holds in \nRollBracket\ of \nRollWindows\ windows. Generalised
connectedness ranges from \rollCgenMin\ to \rollCgenMax\ and the
tier-clustered index from \rollCclustMin\ to \rollCclustMax.
Table~\ref{tab:v4-rolling} contrasts the \nRollCrisis\ windows that
overlap a crisis span by at least one quarter with the \nRollCalm\ that
do not: every index is higher in crises
($\mathcal{C}^{\mathrm{clust}}$ \rollDiffCclust, Welch $p \rollPCclust$),
the own share falls (\rollDiffThetaown), and the increase comes
through the cross-tier contagion share (\rollCalmThetacontag\ to
\rollCrisisThetacontag, $p \rollPThetacontag$) with within-tier
co-movement unchanged (\rollDiffThetacomove). Consecutive rolling
windows share three quarters of their weeks; on each of the
\nRollPhases\ sets of non-overlapping windows the differences keep their
sign, with Welch $p$ between \rollNonPminCclust\ and \rollNonPmaxCclust\
for the clustered index. This is the crisis signature of
Section~\ref{sec:framework} seen from the variance-decomposition side,
and the two sides move together on modularity but not on the spectral
radius: on the shared calendar, Spearman correlations with the NAMH
spectral radius $\rho(A(t))$ are \rhoAllCgenrhoA\ for the generalised
index, \rhoAllCclustrhoA\ for the clustered index and
\rhoAllThetacontagrhoA\ for the contagion share (none significant), and
with modularity $Q(t)$ they are \rhoAllCgenQres\ and
\rhoAllThetacontagQres\ across all windows and \rhoNonCgenQres\ and
\rhoNonThetacontagQres\ on the non-overlapping ones
($p \rhoPNonCgenQres$ and $p \rhoPNonThetacontagQres$). The HWTE network,
estimated at daily frequency from conditional mutual information, and the
weekly variance decomposition, estimated from a linear VAR, agree on when
the system is least modular.

\begin{table}[htbp]
\centering
\small
\caption{Rolling windows: crisis versus calm means, Welch and Wilcoxon
$p$-values.}
\label{tab:v4-rolling}
\begin{tabular}{lcccccrrc}
\toprule
Measure & calm & crisis & mean calm & mean crisis & $\Delta$ & Welch $p$ & Wilcoxon $p$ & non-overlapping: Welch $p$ range \\
\midrule
$\mathcal{C}^{gen}$ & 25 & 19 & 71.7 & 78.7 & $+7.1$ & $<0.001$ & $<0.001$ & [0.010, 0.104] \\
$\mathcal{C}^{clust}$ & 25 & 19 & 62.1 & 69.9 & $+7.7$ & $<0.001$ & $<0.001$ & [0.013, 0.128] \\
$\mathcal{C}^{orth}$ & 25 & 19 & 58.0 & 65.0 & $+7.0$ & $<0.001$ & $<0.001$ & [0.036, 0.206] \\
$\Theta^{own}$ & 25 & 19 & 0.386 & 0.308 & $-0.078$ & $<0.001$ & $<0.001$ & [0.027, 0.130] \\
$\Theta^{comove}$ & 25 & 19 & 0.144 & 0.144 & $+0.000$ & 0.883 & 0.924 & [0.609, 0.933]$^\dagger$ \\
$\Theta^{contag}$ & 25 & 19 & 0.470 & 0.548 & $+0.078$ & $<0.001$ & $<0.001$ & [0.027, 0.139] \\
$\Theta^{comove}-\Theta^{contag}$ & 25 & 19 & -0.326 & -0.404 & $-0.078$ & $<0.001$ & $<0.001$ & [0.030, 0.161] \\
\bottomrule
\end{tabular}
\end{table}

\subsection{Which partition organises connectedness?}
\label{subsec:horse-race}

\begin{figure}[htbp]
\centering
\includegraphics[width=\textwidth]{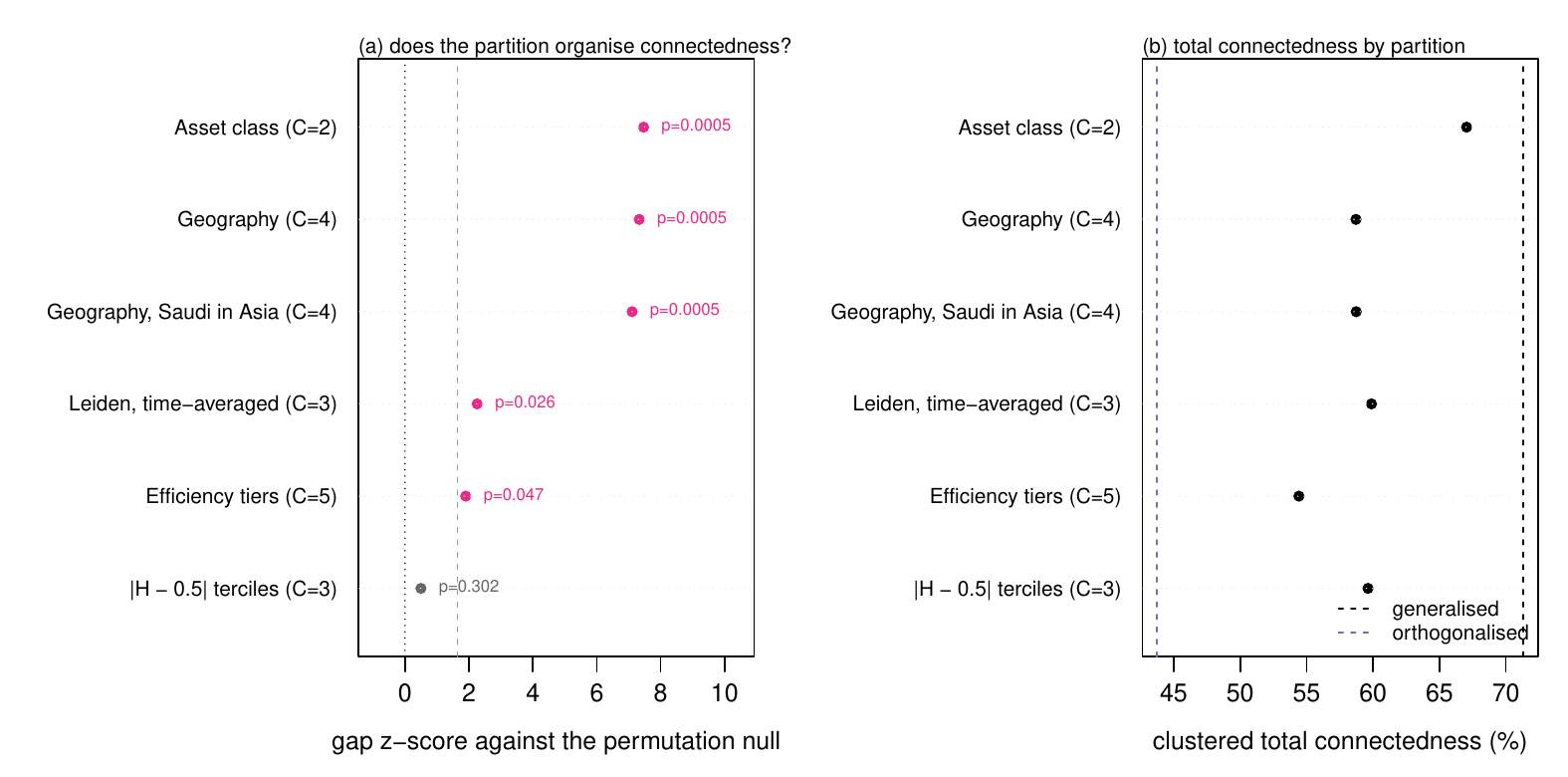}
\caption{Partition horse race: (a) permutation $z$-scores of the blockwise
gap; (b) total connectedness under each partition.}
\label{fig:horse-race}
\end{figure}

The clustered identification takes the partition as given.
\citet{buchwalterDieboldYilmaz2026clustered} assume regions and name
data-driven classification as the open question. The blockwise
permutation test of Proposition~\ref{prop:permutation} lets us ask, for
any candidate partition, whether it separates within-block co-movement
from cross-block contagion more than a random partition of the same block
sizes. We run \nPartitions\ partitions of the same \nAssets\ markets
through identical machinery ($B = \permB$): the efficiency tiers;
univariate terciles of $|\bar H_i - 0.5|$, to see whether the
multivariate clustering adds anything; geography (Americas, Europe and
Middle East, Asia-Pacific, commodities) and the same with Saudi Arabia in
Asia-Pacific; asset
class (equities versus commodities); and the Leiden communities of the
time-averaged magnitude-30 HWTE network. Table~\ref{tab:v4-horse-race}
and Figure~\ref{fig:horse-race} give the result. Geography
($z = \hrZGeography$, $p \hrPGeography$) and asset class
($z = \hrZAssetClass$, $p \hrPAssetClass$) organise the connectedness
network decisively; the Leiden communities of the HWTE network do so
moderately ($p \hrPLeidenTimeAvg$); the efficiency tiers do so only
marginally ($z = \hrZEfficiencyTiers$, $p \hrPEfficiencyTiers$) and the
Hurst terciles not at all ($p \hrPHurstTerciles$). The partitions are nearly independent of
one another (Table~\ref{tab:v4-partition-ari}: ARI between the tiers and
geography is \ariTiersGeo, between the tiers and the Leiden communities
\ariTiersLeiden), so this is not a case of one partition proxying for
another.

\begin{table}[htbp]
\centering
\small
\caption{Partition horse race: total connectedness and decomposition
under each partition's clustered identification (ordering-averaged),
observed gap $\Theta^{comove}-\Theta^{contag}$, permutation $z$-score and
one-sided $p$-value ($B = \permB$, block sizes fixed).}
\label{tab:v4-horse-race}
\begin{tabular}{lclccccccr}
\toprule
Partition & $C$ & sizes & $\mathcal{C}$ & $\Theta^{own}$ & $\Theta^{comove}$ & $\Theta^{contag}$ & gap & $z$ & perm.\ $p$ \\
\midrule
Asset class & 2 & 17/5 & 67.0 & 0.427 & 0.461 & 0.112 & $+0.349$ & 7.46 & $<0.001$ \\
Geography (regions and commodities) & 4 & 5/6/6/5 & 58.7 & 0.418 & 0.227 & 0.356 & $-0.129$ & 7.33 & $<0.001$ \\
Geography, Saudi Arabia in Asia-Pacific & 4 & 5/5/7/5 & 58.7 & 0.409 & 0.232 & 0.359 & $-0.127$ & 7.10 & $<0.001$ \\
Leiden, time-averaged HWTE & 3 & 9/7/6 & 59.9 & 0.379 & 0.224 & 0.397 & $-0.173$ & 2.26 & 0.026 \\
Efficiency tiers (multivariate consensus) & 5 & 4/4/6/7/1 & 54.4 & 0.462 & 0.131 & 0.407 & $-0.276$ & 1.90 & 0.047 \\
$|\bar H-0.5|$ terciles & 3 & 7/8/7 & 59.6 & 0.406 & 0.184 & 0.410 & $-0.227$ & 0.50 & 0.302 \\
\bottomrule
\end{tabular}
\end{table}

\begin{table}[htbp]
\centering
\caption{Adjusted Rand Index between the \nPartitions\ partitions.}
\label{tab:v4-partition-ari}
\begin{tabular}{lcccccc}
\toprule
 & Tiers & $H$ terc. & Geog. & Geog.\ (SA) & Asset & Leiden \\
\midrule
Efficiency tiers (multivariate consensus) & 1.000 & 0.010 & 0.119 & 0.113 & 0.070 & 0.067 \\
$|\bar H-0.5|$ terciles & 0.010 & 1.000 & 0.153 & 0.079 & 0.028 & 0.105 \\
Geography (regions and commodities) & 0.119 & 0.153 & 1.000 & 0.861 & 0.277 & 0.229 \\
Geography, Saudi Arabia in Asia-Pacific & 0.113 & 0.079 & 0.861 & 1.000 & 0.283 & 0.178 \\
Asset class & 0.070 & 0.028 & 0.277 & 0.283 & 1.000 & 0.102 \\
Leiden, time-averaged HWTE & 0.067 & 0.105 & 0.229 & 0.178 & 0.102 & 1.000 \\
\bottomrule
\end{tabular}
\end{table}

\begin{figure}[htbp]
\centering
\includegraphics[width=0.8\textwidth]{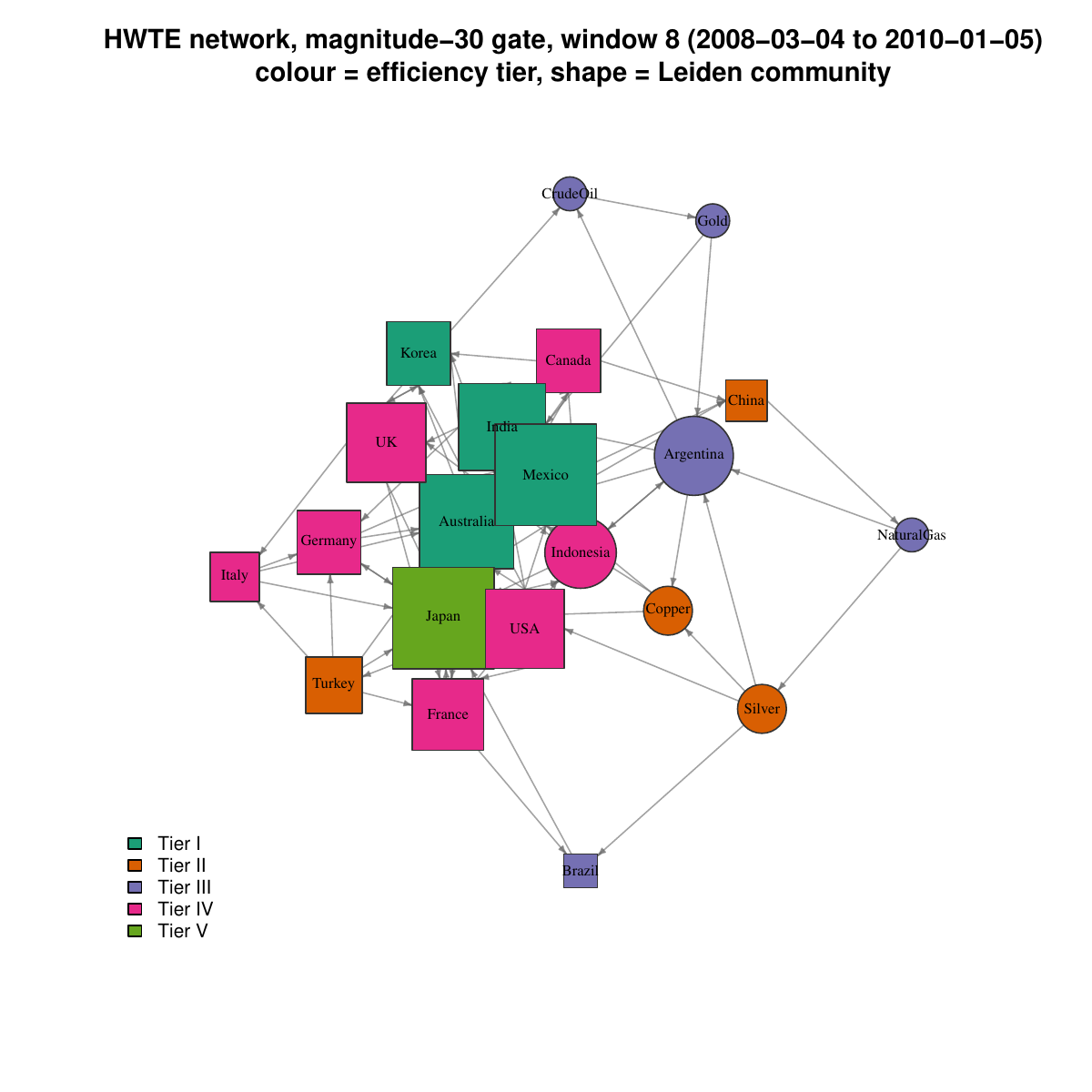}
\caption{HWTE network, magnitude-30 gate, window 8 (the first GFC window). Node
colour: efficiency tier; node shape: that window's Leiden community; node
size: weighted degree.}
\label{fig:network-gfc}
\end{figure}

\begin{figure}[htbp]
\centering
\includegraphics[width=0.7\textwidth]{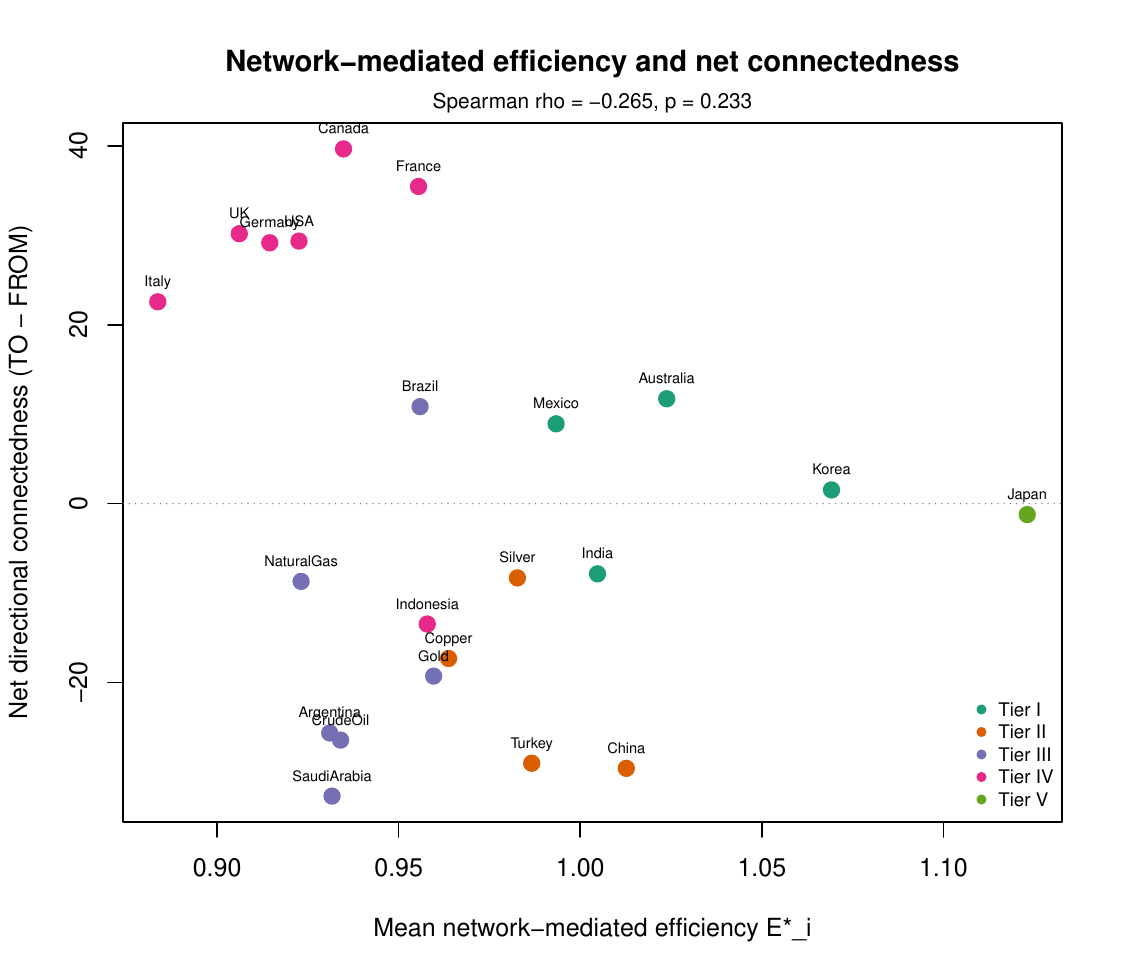}
\caption{Mean network-mediated efficiency $\bar E^*_i$ against generalised
net connectedness, coloured by tier.}
\label{fig:estar-vs-net}
\end{figure}

Two complementary checks point the same way. Running Leiden within each
of the \nWindows\ windows separately, as the canonical pipeline does
(Section~\ref{subsec:community-detection}), the mean ARI between
efficiency tier and that window's communities is \ariObsU\ against a
permutation null with standard deviation \ariNullSdU\ ($p \ariPU$;
Figure~\ref{fig:network-gfc} shows the GFC window). And mean
network-mediated efficiency does not predict net transmission under the
generalised identification (Spearman \spearmanEstarNet,
$p \spearmanEstarNetP$; Figure~\ref{fig:estar-vs-net}). The efficiency
tiers are real, identical under two of the three clustering methods and
interpretable, and they carry a clear
cross-sectional signal in the own / co-movement decomposition; what they
do not do is predict which markets share an information-flow community.
Statistical efficiency and connectedness community are distinct
organising principles of this panel. For the classification question of
\citet{buchwalterDieboldYilmaz2026clustered} the implication is direct:
a data-driven partition should be chosen on the connectedness object
itself (here, geography or asset class, or the network's own
communities), not on a market attribute such as efficiency.

\subsection{Finite-sample behaviour of the permutation test}
\label{subsec:mc-check}

\begin{figure}[htbp]
\centering
\includegraphics[width=0.6\textwidth]{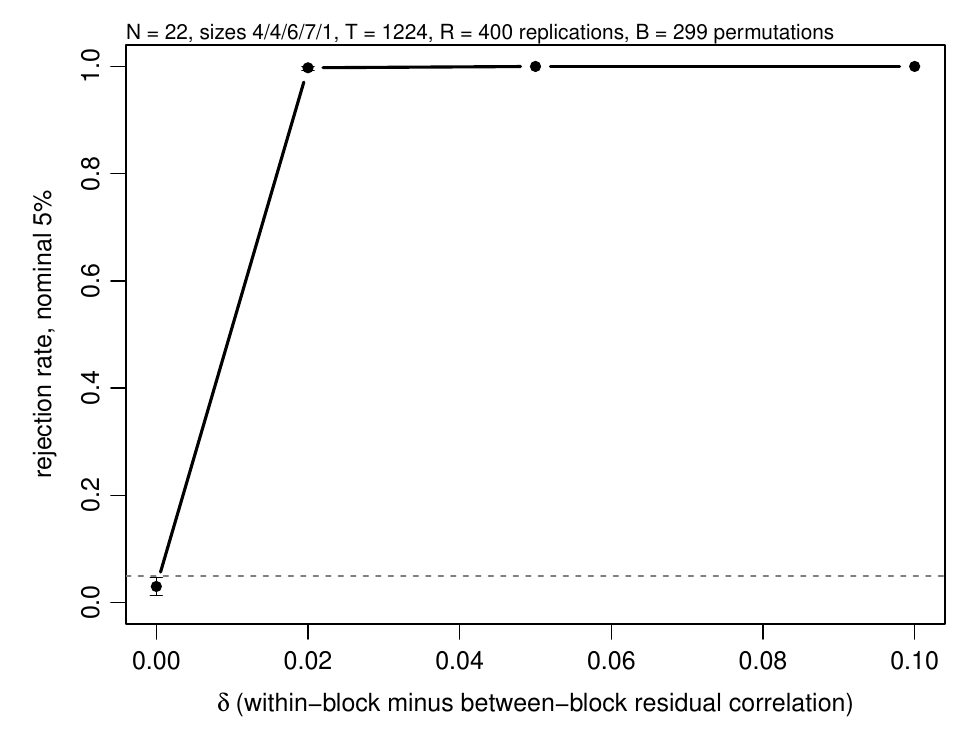}
\caption{Rejection rate of the blockwise permutation test against the
within-block correlation excess $\delta$; bars are $\pm 1.96$ Monte-Carlo
standard errors; dashed line at the nominal 5\% level.}
\label{fig:mc}
\end{figure}

Proposition~\ref{prop:permutation} gives exactness under the
randomisation null; it says nothing about power against a covariance-block
alternative, and the null it controls concerns the labels, not the
data-generating process. We therefore simulate a VAR(1) at the empirical
design ($N = \nAssets$, blocks of $\mcSizes$, $T = \nWeeksFmt$, $\Phi = 0.1\,I$)
with residual correlation $\rho_b = 0.30$ between blocks and
$\rho_b + \delta$ within blocks, and apply the test with the true
partition (\mcR\ replications, \mcB\ permutations each;
Table~\ref{tab:v4-mc}, Figure~\ref{fig:mc}). At $\delta = 0$ the
rejection rate at the 5\% level is \mcSize\ (Monte-Carlo standard error
\mcSizeSe); power reaches \mcPowerTwo\ at $\delta = 0.02$,
\mcPowerFive\ at $\delta = 0.05$ and \mcPowerTen\ at $\delta = 0.10$.
The test is therefore able to detect block structure far weaker than
the regional and asset-class structure it detects in the data, which
sharpens the reading of the marginal efficiency-tier result in
Table~\ref{tab:v4-horse-race}: it reflects a small effect, not a power
failure.

\begin{table}[htbp]
\centering
\caption{Monte-Carlo size and power of the blockwise permutation test at
the empirical design.}
\label{tab:v4-mc}
\begin{tabular}{lccc}
\toprule
$\delta$ & rejection rate, 5\% (MC s.e.) & rejection rate, 10\% & mean gap \\
\midrule
0.00 & 0.030 (0.009) & 0.085 & -0.245 \\
0.02 & 0.998 (0.002) & 0.998 & -0.228 \\
0.05 & 1.000 (0.000) & 1.000 & -0.199 \\
0.10 & 1.000 (0.000) & 1.000 & -0.139 \\
\bottomrule
\end{tabular}
\end{table}

\subsection{Robustness and limitations of the clustered layer}
\label{subsec:eff-robust}

\paragraph{Saudi Arabia and sample length.} Saudi Arabia is in the
connectedness panel over the full sample, although its exchange changed
its weekend in June~2013 and its Hurst exponent is not estimable in
windows \saudiAbsentWindows. Generalised connectedness is
\saudiCgenFullTwo\ with and \saudiCgenFullOne\ without Saudi Arabia on
the full sample, \saudiCgenPreTwo\ and \saudiCgenPreOne\ before the
change, and \saudiCgenPostTwo\ and \saudiCgenPostOne\ after it
(Table~\ref{tab:v4-sa}); net-connectedness rankings have Spearman
correlation \saudiRankFull\ between the panels with and without Saudi
Arabia, and \saudiRankPrePost\ between the samples before and after the
change. Saudi Arabia is the largest net receiver when included (NET
\saudiNetFull). In the NAMH layer, dropping Saudi Arabia from every
window moves the mean spectral radius from \convRhoFull\ to
\convRhoNoSaudi\ and leaves the eigenvector hub unchanged in a fraction
\convHubAgreeNoSaudi\ of windows; setting its missing Hurst exponent to
$0.5$ instead gives \convRhoImp\ and \convHubAgreeImp.

\begin{table}[htbp]
\centering
\caption{Total connectedness with and without Saudi Arabia, by sample.}
\label{tab:v4-sa}
\begin{tabular}{lcccccc}
\toprule
Sample & weeks & $p$ & $\mathcal{C}^{gen}$ & $\mathcal{C}^{clust}$ & $\mathcal{C}^{orth}$ & Saudi NET \\
\midrule
Full sample, $N=22$ & 1224 & 1 & 71.3 & 54.4 & 43.7 & $-32.7$ \\
Full sample, $N=21$ (no Saudi Arabia) & 1224 & 1 & 72.1 & 55.3 & 45.0 &  \\
\midrule
Before 29 June 2013, $N=22$ & 592 & 1 & 74.6 & 58.6 & 48.4 & $-29.0$ \\
Before 29 June 2013, $N=21$ & 592 & 1 & 76.2 & 59.5 & 49.8 &  \\
\midrule
From 29 June 2013, $N=22$ & 632 & 1 & 69.9 & 53.6 & 43.8 & $-21.1$ \\
From 29 June 2013, $N=21$ & 632 & 1 & 69.3 & 53.7 & 44.1 &  \\
\bottomrule
\end{tabular}
\end{table}

\paragraph{Frequency and window.} The connectedness layer runs on weekly
returns and 104-week windows because a \nAssets-variable VAR is not estimable
on the 252-observation windows of the HWTE layer; the two are compared on
a shared calendar, not window for window. Per-window information
criteria would choose three lags on 104 weeks (64 coefficients per
equation) and inflate the level of every index; we hold the order at one
and note the choice.

\paragraph{Method dependence of the tiers.} The tiers depend on the
clustering method (Table~\ref{tab:v4-ari}). The consensus partition is
used precisely so that no single method is treated as authoritative, and
the horse race shows that the univariate Hurst terciles, which depend on
no clustering method, organise connectedness even less than the tiers.

\paragraph{What the tests do and do not establish.} The permutation test
is exact for the randomisation null and, by
Section~\ref{subsec:mc-check}, powerful against covariance-block
alternatives of modest size in a Gaussian VAR(1). It has not been studied
under heavy tails, time-varying volatility or structural breaks, and the
null it controls is conditional on the estimated VAR. The crisis
contrasts of the NAMH tier decomposition rest on \nCrisisWindowsWord\ crisis windows.

\section{Robustness}
\label{sec:robust}

\begin{figure}[!htbp]
\centering
\includegraphics[width=\textwidth]{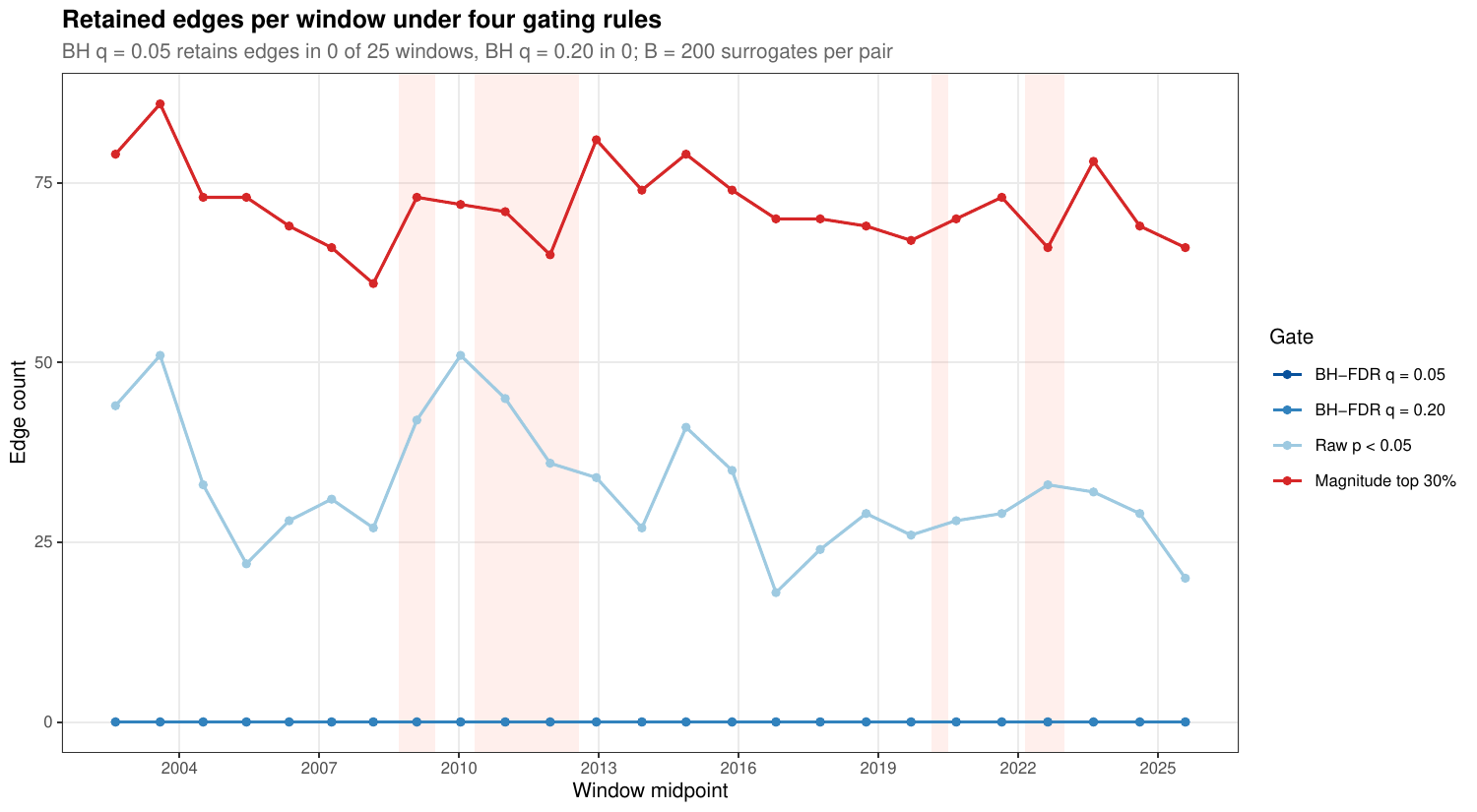}
\caption[Edge counts under four gating modes]{Retained directed-edge
counts per window under the four gating modes evaluated: strict
Benjamini-Hochberg $q < 0.05$, relaxed BH $q < 0.20$, uncorrected
$p < 0.05$, and magnitude top-30\% of positive effective TE. The
maximum number of possible directed edges on the \nAssets-asset network is
$N(N-1) = \nPairs$.}
\label{fig:gating}
\end{figure}

\paragraph{Edge-gating sensitivity.} Figure \ref{fig:gating} reports
the retained-edge counts under the four gating rules evaluated. The
strict Benjamini-Hochberg adjustment at $q < 0.05$ retains zero edges
in \nWindows\ of \nWindows\ windows with $B = \Bhead$ surrogates and
with $B = \Bfull$; the relaxed $q < 0.20$ adjustment retains no edge
with $B = \Bhead$ and, with $B = \Bfull$, a non-zero network in
\nBHwinBfullQTwenty\ windows (\BHwinListBfullQTwenty), with
\nBHedgesBfullQTwenty\ retained directed edges in total. The uncorrected
$p < 0.05$ rule retains \pFiveEdgesMin-\pFiveEdgesMax\ edges per window.
The magnitude top-30\% rule retains \edgesMin-\edgesMax\ edges per
window. The strict-FDR null result is a direct consequence of the
discreteness of the IAAFT surrogate $p$-value: with $B = \Bhead$ shuffles
the floor on the raw $p$ is $1/(B+1) \approx$ \pFloorHead, while the
smallest BH critical value at family size $m = N(N{-}1) = \nPairs$ is
$\alpha/m \approx$ \bhCritMin. The procedure can therefore reject only
if at least $m/((B+1)\alpha)$ pairs attain the floor together:
\bhNeedHeadFive\ pairs at $q = 0.05$ and \bhNeedHeadTwenty\ at $q = 0.20$
with $B = \Bhead$, against \bhNeedFullFive\ and \bhNeedFullTwenty\ with
$B = \Bfull$. The empty strict-FDR networks are thus largely a property of
the surrogate count; we therefore adopt the magnitude-30 rule as the
headline for direct comparability with the prior literature on
threshold-based information-flow networks.

\paragraph{$\varphi$-kernel sensitivity.} We refit the fixed-point
solve with quadratic ($1 - 4(H - 0.5)^2$) and Gaussian
($\exp(-(H - 0.5)^2/(2\sigma^2))$, $\sigma = 0.15$) local-efficiency
kernels on the magnitude-30 HWTE matrices. The panel-mean $\bar E^*$
is $\phiEstarLin$ under linear-absolute (the headline kernel),
$\phiEstarGauss$ under Gaussian, and $\phiEstarQuad$ under quadratic. The
kernel enters $\boldsymbol\varphi$ only, not $A(t)$, so the network
quantities -- the absence of crisis elevation in the spectral radius, the
modularity minimum in \QminWin, the Japan-dominated eigenvector hub -- are
the same under all three kernels.

\paragraph{$\omega$-kernel sensitivity.} We re-fit the HWTE adjacency
with the Gaussian $\omega$-kernel \eqref{eq:omega-gauss},
$\sigma_1 = \sigma_2 = 0.15$, and recompute the eigenvector top-1 hub
per window. Under the Gaussian kernel, Japan tops the eigenvector in
\hubTopGauss\ of \nWindows\ windows (versus \hubTopAll\ under the
exponential kernel), and the two kernels name the same top-1 hub in
\hubAgreeKernels\ of \nWindows\ windows. The Japan-as-eigenvector-hub
finding is therefore robust to the choice of $\omega$-kernel.

\paragraph{Gating-mode sensitivity for hub identity.} Under the
uncorrected $p < 0.05$ gating rule the eigenvector hub identity
disperses considerably: Japan tops only \hubTopPFive\ of \nWindows\
windows, with the hubs scattered across \hubTallyPFive\ (no eigenvector
hub is defined in \nHubNAPFive\ windows). The Japan-as-hub finding is
therefore conditional on the magnitude-30 gating rule. This conditional
reading is appropriate: magnitude-30 retains a stable rich-structure
network (\edgesMin-\edgesMax\ edges per window) in which the Asian
liquidity hub dominates; $p < 0.05$ retains a sparser
($\sim \pFiveEdgesMean$-edge) network in which the hub identity is less
stable.

\paragraph{$\lambda$ leave-one-window-out cross-validation.} For each
$\lambda \in \{1, 2, 5, 10\}$ we evaluate a leave-one-window-out
mean-absolute-deviation score, defined as
$\mathrm{MAE}(\lambda) = W^{-1}\sum_t |\bar E^*_t -
\mathrm{med}_{s \neq t} \bar E^*_s|$. The scores are
$\mathrm{MAE} =$ \lambdaMAE\ for $\lambda = 1, 2, 5, 10$ respectively.
The CV surface is essentially flat (maximum-minimum spread is
$\lambdaMAEspread$, or $\lambdaMAEspreadPct\%$ of the mean), with a
mild preference for larger $\lambda$. The panel-mean spectral radius
$\rho(A)$ \lambdaRhoMonotone\ in $\lambda$ from $\lambdaRhoOne$ at
$\lambda = 1$ to $\lambdaRhoTen$ at $\lambda = 10$, consistent with the
multiplicative-filter interpretation. We retain $\lambda = 5$ for the
headline.

\paragraph{Hurst-estimator triangulation.} We compute three full-
sample Hurst estimates per market: DFA-1 (the rolling-window mean
used as the headline), \citet{geweke1983gph} GPH, and the Sperio
extension of GPH as a local-Whittle proxy. The Spearman rank
correlation across the \nAssets-market panel is
$\rho_S(\mathrm{DFA},\mathrm{GPH}) = \spDFAGPH$,
$\rho_S(\mathrm{DFA},\mathrm{LW}) = \spDFALW$, and
$\rho_S(\mathrm{GPH},\mathrm{LW}) = \spGPHLW$. The mean absolute level
disagreement is $|\mathrm{DFA}-\mathrm{GPH}| = \adDFAGPH$,
$|\mathrm{DFA}-\mathrm{LW}| = \adDFALW$, $|\mathrm{GPH}-\mathrm{LW}| =
\adGPHLW$. GPH and local-Whittle agree closely with each other; DFA-1,
which is a rolling-window mean of within-window scaling exponents
rather than a single full-sample estimate, captures a different
quantity. The implication is that the $\hat H$ panel of Figure
\ref{fig:hurst} should be read as a regime-local quantity rather
than a full-sample long-memory parameter -- the within-window scaling
of integrated returns at scales $s \in [10, 63]$ trading days, which
is precisely the quantity the framework requires.

\paragraph{One-year windows.} We re-estimate the network layer and the
rolling comparison on \annWindows\ non-overlapping windows of one
calendar year each, with the same crisis rule; the ESDC span, longer
than two years, then covers half of no window, and the crisis windows
are \annCrisisList. The GFC year carries the largest spectral radius of
the sample (\annRhoMax, against a median of \annRhoMedian), but the
COVID-19 and Russia-Ukraine years do not, and the crisis mean
(\annRhoCrisis) is not distinguishable from the calm mean
(\annRhoCalm; Wilcoxon $p \annRhoP$). The Spearman correlations of the
three connectedness series with $\rho(A(t))$ lie between \annCrhoMin\
and \annCrhoMax\ ($p \geq \annCrhoPmin$). The absence of a crisis
signature in the spectral radius is therefore not a consequence of the
two-year window.

\section{Concluding remarks}
\label{sec:conclusion}

The Network Adaptive Market Hypothesis provides a quantitatively
tractable, empirically observable parametrisation of the Grossman-
Stiglitz interior that integrates long-memory econometrics, information
theory, and network analysis into a single estimable system. The
framework's central object -- the Hurst-Weighted Transfer Entropy
adjacency $A(t)$ -- is a directionally-asymmetric, fractal-aware,
surrogate-bias-corrected measure of cross-market information flow whose
spectral and modular properties supply a falsifiable diagnostic for
the network signature of systemic stress.

Three substantive findings emerge from the empirical analysis. First,
the eigenvector hub of the estimated information network on the
G20-plus-commodities panel is \emph{Japan} in \hubTopAll\ of \nWindows\
two-year windows, while the U.S.\ functions as a \emph{broadcaster}
(rank 1 on out-strength in \usaOutOneN\ of \nWindows\ windows, all of
them crisis windows) and a \emph{cross-community bridge} in crises (rank
\usaBetwBest\ on weighted-directed betweenness in \usaBetwBestList).
Second, the dense-network amplification phase transition of
\citet{acemogluOzdaglarTahbazSalehi2015systemic} is \emph{not} detected
as a crisis signature -- the spectral radius is no higher in the GFC,
ESDC, COVID-19 or Russia-Ukraine windows than in calm ones, and
modularity is lower in crises only weakly -- suggesting that the
network-mediated amplification mechanism, if present, is not visible at
the two-year resolution of the windows, nor at a one-year resolution
outside the GFC. Third, the leave-one-out
decomposition of the spectral radius shows that the crisis-window core
is recomposed toward \rhoUpFiveNames, while contributions from
\rhoDownFiveNames\ fall during the same windows.

A fourth finding comes from reading the framework through a partition of
the panel into efficiency tiers and placing it beside the
variance-decomposition connectedness apparatus. The two lenses agree on
the dynamics: clustered connectedness sits between its generalised and
orthogonalised polar cases in all \nRollWindows\ rolling windows, its
cross-cluster share is higher in crisis windows, and both move inversely
with NAMH modularity (Spearman \rhoNonCgenQres\ and \rhoNonThetacontagQres\
on the non-overlapping windows), although not with the NAMH spectral
radius (\rhoAllCclustrhoA\ and \rhoAllThetacontagrhoA). They also agree
on the cross-section: the middle efficiency tier, Tier~\idioTierVD, is the
most idiosyncratic, carrying the highest own share under the clustered
variance decomposition (\ownIII\ against \ownI\ for the most efficient
tier) and the highest own share of its own equilibrium efficiency under
the Neumann accounting of Proposition~\ref{prop:neumann-tiers}. What the
tiers do only marginally is organise the connectedness network: the
blockwise permutation test of Proposition~\ref{prop:permutation} rejects
a random partition of the same shape for the efficiency tiers only at
the margin ($p \hrPEfficiencyTiers$), while it rejects decisively for
geography ($p \hrPGeography$) and asset class ($p \hrPAssetClass$). Statistical
efficiency and information-flow community are distinct organising
principles of the panel, which is itself informative for the data-driven
classification problem that \citet{buchwalterDieboldYilmaz2026clustered}
pose.

Three avenues for future work are immediate. First, a structural
microfoundation of HWTE as the equilibrium statistic of a multi-asset
strategic-trading model in the tradition of \citet{caballeKrishnan1994}
and \citet{bernhardtTaub2008} would convert the metaphor of cross-
market price discovery into a derivation from primitives. Second,
embedding the NAMH in a general-equilibrium framework with
heterogeneous-agent New Keynesian dynamics or in an agent-based
computational model with explicit replicator dynamics on a
fundamentalist-chartist-noise-trader population would let the
adjacency $A(t)$ itself emerge endogenously rather than be estimated.
Third, applying the pipeline to higher-frequency intraday data, to
portfolios beyond the G20-plus-commodities universe, and to non-price
information channels -- news flow, central-bank communication,
search-query intensities -- would extend the diagnostic to new asset
classes and information modalities.

The full empirical pipeline -- with the KSG / Frenzel-Pompe transfer-
entropy estimator, IAAFT surrogates with Marschinski-Kantz effective
TE, Benjamini-Hochberg edge gating, Leiden community detection,
canonical-form Katz on the dominant strongly-connected component, and
spectral-radius-safe NAMH solve -- is released as the open-source R
package \texttt{namh}. The clustered-connectedness layer of Section
\ref{sec:efficiency-clustering} and every table and
figure of that section regenerate from the self-contained replication
materials described in Appendix~\ref{app:eff-repro}.

\appendix

\section{Proofs for Section~\ref{subsec:theory-clustered}}
\label{app:proofs-clustered}

Throughout, $[N] = \{1,\ldots,N\}$, $\pi:[N]\to[C]$ is a partition, and
for $i\in[N]$ write $S_i = \{j : \pi(j) = \pi(i)\}$ for $i$'s cluster.

\begin{proof}[Proof of Proposition~\ref{prop:partition-unity}]
Split $\sum_{j\in[N]}\theta_{ij}$ over $S_i$ and its complement, then
split the sum over $S_i$ into $j = i$ and $j \in S_i\setminus\{i\}$:
\[
1 = \sum_{j}\theta_{ij}
  = \theta_{ii} + \sum_{j\in S_i\setminus\{i\}}\theta_{ij} + \sum_{j\notin S_i}\theta_{ij}
  = \theta^{\mathrm{own}}_i + \theta^{\mathrm{within}}_i + \sum_{k\neq\pi(i)}\theta^{\mathrm{cross}}_{i\leftarrow k},
\]
where the last equality regroups $[N]\setminus S_i$ by cluster. Each term
is a sum of non-negative entries, hence in $[0,1]$ by the identity.
Averaging over $i$ with $\pi(i) = c$ and dividing by $N_c$ gives the
cluster identity; averaging over $c$ gives the system identity.
\end{proof}

\begin{proof}[Proof of Proposition~\ref{prop:refinement}]
If $\pi'$ refines $\pi$ then $S'_i\setminus\{i\} \subseteq
S_i\setminus\{i\}$, so $\theta^{\mathrm{within}}_i(\pi')$ is a sub-sum of
non-negative terms of $\theta^{\mathrm{within}}_i(\pi)$. The cross
inequality follows from Proposition~\ref{prop:partition-unity} because the
own share does not depend on the partition. For $C = 1$ the complement of
$S_i$ is empty; for $C = N$, $S_i\setminus\{i\}$ is empty.
\end{proof}

\begin{proof}[Proof of Proposition~\ref{prop:block-orth}]
(i) \emph{Existence.} Let $\mathbf{u} = (\mathbf{u}_1,\ldots,\mathbf{u}_C)$
be the reduced-form residual in block order and define
$\boldsymbol\epsilon_1 = \mathbf{u}_1$, $\boldsymbol\epsilon_c = \mathbf{u}_c -
B_c\mathbf{u}_{<c}$ for $c\geq 2$, i.e.\ $\boldsymbol\epsilon = Q_C^{-1}\mathbf{u}$
with $Q_C^{-1}$ unit lower block-triangular and block row $c$ equal to
$(-B_c, I)$. Since $B_c$ is the population regression coefficient of
$\mathbf{u}_c$ on $\mathbf{u}_{<c}$,
$\operatorname{Cov}(\boldsymbol\epsilon_c,\mathbf{u}_{<c}) = \Sigma_{c,<c} -
B_c\Sigma_{<c,<c} = 0$. Every $\boldsymbol\epsilon_l$ with $l < c$ is a
linear function of $\mathbf{u}_{<c}$, so
$\operatorname{Cov}(\boldsymbol\epsilon_c,\boldsymbol\epsilon_l) = 0$ for
$l\neq c$ and $\Omega_C = \operatorname{Var}(\boldsymbol\epsilon)$ is
block-diagonal with $\Omega_{cc} = \Sigma_{cc} - B_c\Sigma_{<c,c}$, the
Schur complement of $\Sigma_{<c,<c}$ in the leading block of $\Sigma$
through $c$. Unit block-triangularity makes $Q_C^{-1}$, hence $Q_C$,
invertible. \emph{Uniqueness.} Let $P$ be unit lower block-triangular with
$P\Sigma P^\top$ block-diagonal, and let $P_c = (P_{c,<c}, I)$ be its
block row $c$. Block-diagonality means $P_c\mathbf{u}$ is uncorrelated
with $P_l\mathbf{u}$ for all $l < c$. The leading principal block of $P$
through $c-1$ is unit lower block-triangular, hence invertible, so
$\operatorname{span}\{P_l\mathbf{u} : l<c\} = \operatorname{span}\{\mathbf{u}_{<c}\}$
and $P_c\mathbf{u} = \mathbf{u}_c + P_{c,<c}\mathbf{u}_{<c}$ is uncorrelated
with $\mathbf{u}_{<c}$. Thus $\Sigma_{c,<c} + P_{c,<c}\Sigma_{<c,<c} = 0$,
and since $\Sigma_{<c,<c}\succ 0$, $P_{c,<c} = -B_c$. For $C = 3$ with
$B_3 = (\Sigma_{31},\Sigma_{32})\bigl(\begin{smallmatrix}\Sigma_{11}&\Sigma_{12}\\
\Sigma_{21}&\Sigma_{22}\end{smallmatrix}\bigr)^{-1}$ this is equation (7)
of \citet{buchwalterDieboldYilmaz2026clustered}; their Appendix~A derives
the same object by sequential projection.

(ii) For $C = 1$ there are no earlier blocks, so $Q_1^{-1} = I$ and
$\Omega_1 = \Sigma$; substituting into \eqref{eq:clustered-fevd} gives
the generalised decomposition $\sigma_{jj}^{-1}\sum_h(e_i^\top A_h\Sigma
e_j)^2$ of \citet{pesaranShin1998}. For $C = N$ every block is a
singleton, $Q_N^{-1}$ is unit lower triangular and $\Omega_N = D$ is
diagonal with positive entries, so $\Sigma = Q_N D Q_N^\top$ is the
$LDL^\top$ factorisation and $M = Q_N D^{1/2}$ is the (unique) lower
Cholesky factor. Then $Q_N\Omega_N e_j/\sqrt{\omega_{N,jj}} = Q_N D e_j /
\sqrt{d_j} = M e_j$ and \eqref{eq:clustered-fevd} reduces to
$\sum_h(e_i^\top A_h M e_j)^2$, the orthogonalised decomposition.

(iii) Let $\Pi$ be a permutation matrix that permutes markets within
blocks only, and write $\tilde\Sigma = \Pi\Sigma\Pi^\top$, $\tilde A_h =
\Pi A_h\Pi^\top$. Because $\Pi$ maps each block onto itself, the block
regression coefficients transform as $\tilde B_c = \Pi_c B_c
\Pi_{<c}^\top$, so $\tilde Q_C^{-1} = \Pi Q_C^{-1}\Pi^\top$ and
$\tilde\Omega_C = \Pi\Omega_C\Pi^\top$. Substituting into
\eqref{eq:clustered-fevd} and using $\Pi^\top e_{\Pi(j)} = e_j$ gives
$\tilde\theta^{C}_{\Pi(i)\Pi(j)} = \theta^{C}_{ij}$ before and after row
normalisation. A permutation that moves a market across blocks changes the
conditioning set $\mathbf{u}_{<c}$ of at least one block and hence $B_c$;
Section~\ref{subsec:clustered-dy} reports the resulting spread across the
$C!$ block orderings.
\end{proof}

\begin{proof}[Proof of Proposition~\ref{prop:neumann-tiers}]
By Theorem~\ref{thm:exist-unique}, $\rho(A) < 1$ gives $M = \sum_{k\geq
0}A^k$ with the series absolutely convergent. Each $A^k \geq 0$ entrywise
because products and sums of non-negative matrices are non-negative, so
$M \geq 0$ and, for every $K$, $M_K =
\sum_{k<K}A^k \geq 0$. Since $A^0 = I$, $(M_K\mathbf{b})_i = b_i +
\sum_{k=1}^{K-1}(A^k\mathbf{b})_i \geq b_i$ for $K\geq 1$; letting
$K\to\infty$ gives $E^*_i \geq
b_i$. If $E^*_i > 0$, the row $s_{ij} = M_{ij}b_j/E^*_i$ is non-negative
and sums to $\sum_j M_{ij}b_j / E^*_i = 1$, so $[s_{ij}]$ is a share
matrix and (i) is
Proposition~\ref{prop:partition-unity} applied to it. For (ii),
$\mathbf{E}^*$ is linear
in $\mathbf{b}$ with $d\mathbf{E}^* = M\,d\mathbf{b}$; if $d\mathbf{b}\geq
0$ is supported on tier $k$ then $dE^*_i = \sum_{\pi(j)=k}M_{ij}db_j \geq
0$, and the same argument applies to each partial sum $M_K$. For (iii),
$\Psi_{ck} \geq 0$ from $M\geq 0$, and
$\Psi_{cc} \geq N_c^{-1}\sum_{\pi(i)=c}M_{ii} \geq 1$ because $M_{ii} =
1 + \sum_{k\geq 1}(A^k)_{ii} \geq 1$.
\end{proof}

\begin{proof}[Proof of Proposition~\ref{prop:permutation}]
Conditional on the data, $T(\pi)$ is a deterministic function of the
block assignment. Under the randomisation hypothesis the assignment
$\pi$ has the same distribution as $g\pi$ for every $g\in\mathcal{G}$, so
$T(\pi)$ is a uniform draw from the orbit $\{T(g\pi): g\in\mathcal{G}\}$
(counted with multiplicity). The Monte-Carlo $p$-value with $B$
independent uniform draws from $\mathcal{G}$ and the $+1$ convention is
then the $p$-value of a randomisation test in the sense of
\citet[Section~15.2]{lehmannRomano2005testing}, and Theorem~15.2.1 there
gives $\Pr_{H_0}(\hat p \leq \alpha)\leq\alpha$; the inclusion of the
observed statistic in the reference set is what makes the bound hold for
finite $B$ \citep{phipsonSmyth2010permutation}. Averaging over block
orderings inside $T$ does not affect the argument, since the average is
itself a fixed function of the assignment.
\end{proof}

\section{Replication and numerical checks}
\label{app:eff-repro}

\paragraph{Materials.} The results of
Section~\ref{sec:efficiency-clustering} are computed from three inputs
alone: the daily return panel of Section~\ref{sec:data}, the
twenty-five-window estimation of the efficiency network reported in
Section~\ref{sec:results}, and the diagnostic series that accompany it.
The replication materials, which extend the \texttt{namh} package
described in Section~\ref{sec:data}, reproduce every table, figure and
reported quantity of that section from those inputs under a fixed random
seed.

\paragraph{Reproduction of the estimated network.} Two checks precede the
new estimates. First, the effective transfer-entropy and $p$-value
matrices are recomputed in every window from the raw transfer entropies
and the stored surrogate draws, at both surrogate counts; all
twenty-five windows agree exactly. Second, the adjacency $A(t) = \omega(H(t);
\lambda=5)\circ\mathrm{TE}^{\mathrm{eff}}(t)\circ\mathbb{1}\{\text{gate}\}$
and the fixed point $\mathbf{E}^*(t) = (I - A(t))^{-1}\boldsymbol\varphi(t)$
are recomputed from primitives and compared with the estimated
$\mathbf{E}^*(t)$ and $\rho(A(t))$; the largest absolute discrepancy
across all windows and markets is \decompMaxDE. The comparison is made
against the magnitude-gated quantities, which are the ones reported in
Section~\ref{sec:results}.

\paragraph{Numerical checks on the identification.} The
block-orthogonalised decomposition was checked on synthetic covariance
matrices against the three-block closed form of
\citet[eq.~7]{buchwalterDieboldYilmaz2026clustered}, for
block-diagonality of $\Omega_C$, for the row-sum identity of
Definition~\ref{def:share-matrix}, and for the two polar cases of
Proposition~\ref{prop:block-orth}(ii). All four hold as exact identities
up to floating-point rounding.

\paragraph{Computational notes.} Full-sample VAR order is chosen by AIC
over $p\leq 3$ on $T = \nWeeksFmt$ weeks and selects $p = \varLag$. In the rolling
analysis the order is held at $p = \varLag$: per-window AIC selects $p = 3$ in
every one of the \nRollWindows\ 104-week windows, which is $67$ coefficients
per equation on $104$ observations; we do not report that series.
Cluster-ordering averages enumerate all $C!$ orderings ($C \leq 5$ in every
partition used). The Monte-Carlo study of
Section~\ref{subsec:mc-check} and the bootstrap of
Table~\ref{tab:v4-bootstrap} draw their seeds per replication, so the
results do not depend on how the computation is distributed.

\bibliographystyle{plainnat}
\section*{Acknowledgements}
The authors thank the Google Cloud Education Programs for a research credits award under the
Google Cloud Research Credits Programme (credit reference wilsonjessica-485604931), which funds
the cloud infrastructure, AI services, and data-warehouse layer of the deployed platform. The
computational infrastructure was procured under Institute Seed Grant SP128 (Innovative Approaches
for Analyzing Financial Market Dynamics: Computational Financial Analysis with Wavelets, Networks,
and Statistical Learning Powered by High-Performance Computing), funded by the Indian Institute of
Technology Bhubaneswar (Sponsored Research and Industrial Consultancy, sanction dated 18 November
2023).

\end{document}